\documentclass[fleqn,usenatbib]{mnras}

\usepackage{CJKutf8}
\newcommand{\CJKname}{\begin{CJK*}{UTF8}{bkai}黃鎔鈞\end{CJK*}} % or gkai/bsmi/bkai
\usepackage{newtxtext,newtxmath}
\usepackage[T1]{fontenc}

\DeclareRobustCommand{\VAN}[3]{#2}
\let\VANthebibliography\thebibliography
\def\thebibliography{\DeclareRobustCommand{\VAN}[3]{##3}\VANthebibliography}

\usepackage{graphicx}	% Including figure files
\usepackage{amsmath}	% Advanced maths commands

\usepackage{multirow}
\usepackage{diagbox}
\usepackage
{caption}
\usepackage{subcaption}

\usepackage{orcidlink}
\usepackage{xcolor}

\usepackage{hyperref}
\usepackage{appendix}

\usepackage{tabularx}  

\newcommand{\citeg}[1]{\citep[e.g.,][]{#1}}

\newcommand{\aref}[1]{\hyperref[#1]{Appendix~\ref{#1}}}
\defcitealias{vijayan24}{QED I}
\defcitealias{huang25}{QED II}
\defcitealias{vijayan25}{QED III}
\newcommand{\halpha}{H$\alpha$}
\newcommand{\kmps}{km s$^{-1}$}

\newcommand{\qediii}{\citetalias{vijayan25}}
\newcommand{\fid}{$\Sigma$13-Z1-H150}
\newcommand{\subsol}{$\Sigma$13-Z0.2-H150}
\newcommand{\outgal}{$\Sigma$2.5-Z1-H1500}

\title[QED IV - \halpha\ outflow limitations]{\textsc{Quokka}-based understanding of outflows (QED) - IV. Limitations of \halpha\ as an outflow diagnostic}

\author[R. Huang et al.]{Rongjun Huang (\CJKname)\orcidlink{0000-0002-6646-8365}$^{1,2}$\thanks{E-mail: U6569836@anu.edu.au, Astro@Rongjun-Huang.com}, 
Aditi Vijayan\orcidlink{0000-0002-7714-2379}$^{1}$\thanks{E-mail: Aditi.Vijayan@anu.edu.au}, 
Mark R. Krumholz\orcidlink{0000-0003-3893-854X}$^{1}$\thanks{E-mail: Mark.Krumholz@anu.edu.au} 
\\
$^{1}$Research School of Astronomy and Astrophysics, Australian National University, Cotter Road, Weston Creek, ACT 2611, Australia\\
$^{2}$International Centre for Radio Astronomy Research, University of Western Australia, Crawley, WA 6009, Australia\\
}

\date{Accepted XXX. Received YYY; in original form ZZZ}

\pubyear{2025}

\begin{document}
\label{firstpage}
\pagerange{\pageref{firstpage}--\pageref{lastpage}}
\maketitle

\begin{abstract}
The presence of broad wings in the \halpha\ line is commonly used as a diagnostic of the presence and properties of galactic winds from star-forming galaxies. However, the accuracy of this approach has not been subjected to extensive testing. In this paper, we use high-resolution simulations of galactic wind launching to calibrate the extent to which broad \halpha\ wings can be used to infer the properties of galactic outflows. For this purpose, we analyse a series of high-resolution wind simulations from the QED suite spanning two orders of magnitude in star formation surface density ($\Sigma_\mathrm{SFR}$). We show that the broad component of \halpha\ emission correlates well with the wind mass flux at heights $\sim1$ kpc above the galactic plane, but that the correlation is poor at larger distances from the plane, and that even at 1 kpc the relationship between mass flux and surface brightness of broad \halpha\ is significantly sub-linear. The sub-linear scaling suggests that the electron column density in the wind increases systematically with outflow strength, and that the conventional assumption of constant electron density in the wind leads to a systematic overestimate of how steeply mass loading factors depend on $\Sigma_\mathrm{SFR}$. We provide empirical scaling relations that observers can apply to correct for this effect when converting \halpha\ measurements to mass outflow rates. Finally, we use synthetic observations of the density-diagnostic $[\ion{S}{ii}]\,\lambda\lambda6716,6731$ doublet to show that using this diagnostic only slightly improves estimates of wind outflow rates compared to the naive assumption of constant electron density, and performs significantly worse than the empirical correlation we provide.
\end{abstract}


%%%%%%%%%%%%%%%%%%%%%%%%%%%%%%%%%%%%%%%%%%%%%%%%%%

%%%%%%%%%%%%%%%%% BODY OF PAPER %%%%%%%%%%%%%%%%%%

%=======================================================================
\section{Introduction}
\label{Sec:intro}

% \textbf{Observational motivation}  
%         \begin{itemize}
%             \item Broad ($|v|\gtrsim 50 kms^{-1}$) \halpha\ wings are routinely detected in IFU surveys such as SAMI \citep{zovaro24} and MaNGA, and are interpreted as tracers of galactic winds.  
%             \item Converting wing luminosity to mass–outflow rate usually assumes a fixed electron density, e.g. $n_e = 100 cm^{-3}$ in the DUVET survey \citep{reichardtchu22}, and similar choices in high‐$z$ studies \citep{genzel14}.   
%         \end{itemize}

Star formation processes occurring in the disc of a galaxy drive large-scale outflows of multiphase gas that are observable in wavebands from X-ray \citep{lopez20, Lopez23a} to UV \citep{chisholm18, Xu22a} to optical \citep{Wood+15,  reichardtchu22, Hamel-Bravo24a, zovaro24} to infrared \citep{fisher25} to radio \citep{leroy15, martini18}. They form a part of the larger baryonic cycle in galaxies and are responsible for transferring mass, momentum, energy, and metals between a galaxy and its surrounding circumgalactic medium (CGM). Because of their importance to this cycle, making accurate measurements of the mass flux carried by these outflows has been a major focus of both observational and theoretical studies. While the hot ($\gtrsim 10^6$ K) supernova-driven phase is believed to carry most of the outflow energy and metals, the majority of the outflow mass is thought to be carried by the cooler phases: warm ionised gas at $\sim 10^4$ K, atomic gas at $\approx 200 - 5000$ K, and molecular gas at $\lesssim 50$ K \citep[e.g.,][]{kim18, kim20,  veilleux20, Yuan23a, Thompson24a, vijayan24, vijayan25}. In emission, the latter two phases are accessible primarily in the radio, while for the former, the most prominent emission tracer is the \halpha\ line. However, converting \halpha\ emission into a mass flux requires an estimate of the electron density, and different observational diagnostics can yield systematically different $n_e$ values for the same galaxies \citeg{xu25}. This line -- and the extent to which we can use it to diagnose mass outflow rates in galactic winds -- is the primary focus of this paper.

The first challenge for studies of outflows in \halpha\ is to separate emission from the outflow from the (usually much brighter) emission from photoionised gas in the galactic disc. When the target is close to edge-on and spatially resolved, as is the case for the prominent nearby starburst M82 \citep[e.g.,][]{Shopbell98a, Martin98a}, this separation is easily accomplished simply by targeting parts of the emission off the disc. However, in less well-resolved or less favourably-aligned systems, separation is generally accomplished spectroscopically. In these systems, outflowing gas is detected as a broad, faint component of the \halpha\ line ($|v|\gtrsim 50$ km s$^{-1}$) that sits on top of the narrow, bright component associated with ionisation of gas from stars in the disc of the galaxy.

Large integral field unit (IFU) spectroscopy surveys have established that such broad components are common in star-forming systems, both locally and at cosmological distances. For the latter, surveys such as the Spectroscopic Imaging survey in the Near-infrared with SINFONI \citep[SINS;][]{Genzel+11, newman12}, the KMOS Redshift One Spectroscopic Survey \citep[KROSS;][]{Swinbank19a}, and the KMOS$^{3\mathrm{D}}$ survey \citep{Schreiber+19} have revealed ubiquitous broad \halpha\ emission from galaxies at cosmic noon, and authors have estimated the outflow rates and mass loading factors associated with these outflows using a variety of methods. Locally, the Sydney-Australian-Astronomical-Observatory Multi-object Integral-Field Spectrograph (SAMI) galaxy survey roughly 28\% of galaxies have \halpha\ spectra that are best fit as the sum of narrow and broad components rather than as single-component \citep{oh24}. The incidence of such multi-component line profiles correlates strongly with star formation intensity -- galaxies with high star formation rate surface density ($\Sigma_{\rm SFR}$) are much more likely to exhibit broad-wing features \citep{zovaro24}. Likewise, a systematic study in the Mapping Nearby Galaxies at APO (MaNGA) survey found broad \halpha\ wings (typical full width at half maximum of a few hundred km s$^{-1}$) in about 7\% of local star-forming galaxies, which are driven by stellar feedback or Active Galactic Nucleus (AGN) activity \citep{rodriguez19}. The Calar Alto Legacy Integral Field Area (CALIFA) survey also reports widespread extraplanar ionised gas and outflow candidates in normal discs \citep[e.g.][]{lopezcoba19}. Hence, these studies establish the empirical basis for using \halpha\ emission wings to study galactic winds.

However, fundamental uncertainties remain about what \halpha\ broad components actually trace. For example, the nature of the outflow motion, the temperature range of the emitting gas, the gas density, and the spatial origin of the emission are all uncertain. This makes it challenging to derive outflow rates -- the quantity in which we are ultimately most interested -- from \halpha\ data, and often forces studies to rely on assumptions of unknown accuracy about the physical properties of the emitting gas. For example, observational studies often convert the \halpha\ broad component luminosity into a mass outflow rate by assuming the ionised gas is radiating under case B recombination with some characteristic electron density ($n_e$). In practice, such studies usually just adopt a fixed density due to the lack of direct constraints, typically $n_e \sim 100 - 200$ $\rm{cm^{-3}}$. For instance, the Deep near-UV observations of Entrained gas in Turbulent galaxies (DUVET) survey of nearby starbursts assumed $n_e = 100$ $\rm{cm^{-3}}$ when estimating the mass carried by the \halpha-emitting outflows \citep{reichardtchu22}. At high redshift, similar assumptions are made: \citet{genzel14} studied massive $z\sim2$ star-forming galaxies by assuming an outflow electron density of $n_e = 80$ $\rm{cm^{-3}}$, while a survey of ultra-luminous starbursts by \citet{fiore17} adopted $n_e = 200$ $\rm{cm^{-3}}$ uniformly for the outflowing gas. 

This ``fixed-density'' approach is widespread, but it carries considerable uncertainty: the derived outflow masses scale inversely with the assumed $n_e$, so a misestimate can directly bias the inferred mass outflow rates by up to a factor of 3 \citep{reichardtchu22}. Recent studies indeed suggest that a single canonical density is overly simplistic, since the outflowing gas can have differing values of $n_e$ even from one region to another within an individual galaxy, let alone between galaxies. For example, in local luminous infrared galaxies, \citet{fluetsch21} use density-sensitive line ratios to estimate that the ionised outflow phase has an average density $n_e = 500$ $\rm{cm^{-3}}$, about three times higher than is typical of \ion{H}{ii} regions in the corresponding galactic discs. In the local starburst NGC~4383, \citet{watts24} report an upper limit of a few $100$ ${\rm cm^{-3}}$ on $n_e$ in the bright inner 300 pc region, but find that the density decreases to $\sim1$ ${\rm cm^{-3}}$ at a distance of 6 kpc from the nucleus. In general, a calibration of $n_e$ is needed when interpreting \halpha\ wings and galactic outflows. 

% \textbf{Theory / simulation gap}  
%         \begin{itemize}
%             \item We don't believe electron density is constant across outflows with different strength. And we want to test how bad it is for constant $n_e$ assumption. 
%             \item Our simulation data (the QEDIII suite; \citealt{vijayan25}) resolve the warm phase ($10^{4}\le T/\mathrm{K}\le3\times10^{4}$) and directly output the outflow properties, e.g., ...  
%             \item We analyse three runs ($\rm \Sigma13-Z1-H150$, R8-$0.2Z_\odot$, $\rm \Sigma2.5-Z1-H1500$) spanning two orders of magnitude in $\Sigma_{\rm SFR}$ and exhibiting steady warm outflows.  
%         \end{itemize}

% Our goal in this paper is to use recent high-resolution simulations of galactic outflows that resolve their multiphase structure to provide the required calibration -- or, at a more basic level, to determine whether such a calibration is even possible. 
Our goal in this paper is to use recent high-resolution simulations of galactic outflows that resolve their multiphase structure to test common observational assumptions used to infer mass outflow rates from the \halpha\ line, and thereby to provide a numerical calibration where appropriate. In particular, we seek to explore 1) how well \halpha\ wings trace galactic outflows, and 2) how much error we incur by using the constant $n_e$ assumption to infer outflow rates? In doing so, we consider both (i) \halpha-based estimators that adopt a fixed $n_e$ (or fixed electron column density) and (ii) estimators that infer $n_e$ from nebular line ratios, focusing in particular on the density-sensitive [S\,\textsc{ii}] $\lambda\lambda6716,6731$ doublet. Addressing these questions requires knowing the actual density and ionisation state in outflows driven over a range of galactic environments, something that is uniquely measurable in simulations that reach the parsec‐scale resolution required to capture the warm ($T\sim10^{4}$K) ionised phase that emits \halpha\ and separate it from both the hot ($\gtrsim 10^6$ K) and cooler neutral phases. For this purpose, we use the QED simulation suite\footnote{\textsc{QUOKKA}-based Understanding of Outflows Derived from Extensive, Repeated, Accurate, Thorough, Demanding, Expensive, Memory-consuming, Ongoing Numerical Simulations of Transport, Removal, Accretion, Nucleosynthesis, Deposition, and Uplifting of Metals (QUOD ERAT DEMONSTRANDUM, or QED).} first described by \citet[hereafter \citetalias{vijayan24}]{vijayan24} and further studied by \citet[hereafter \citetalias{huang25}]{huang25} and \citet[hereafter \citetalias{vijayan25}]{vijayan25}. QED is a set of three–dimensional, GPU‐accelerated radiation–hydrodynamic simulations run with the adaptive‐mesh code \textsc{Quokka} \citep{wibking22,He24a} that adopt a `tall‐box’ geometry (1 kpc $\times$ 1 kpc in the disc plane, extending to $\pm4$ or 8 kpc vertically for most simulations) and feature a uniform cell size of 2 or 4 pc throughout the domain. Data from these simulations provide spatially and temporally resolved maps of outflow mass, velocity, and electron density against which we can calibrate the \halpha\ broad component diagnostic investigated here.

This paper is organized as follows: \autoref{sec:data-methods} presents the simulation data and methods; \autoref{sec:results} presents the correlations and the electron–column diagnostics; \autoref{sec:disscusion} discusses implications and \autoref{sec:conclusion} summarizes our findings.

%=======================================================================
\section{Data \& Methods}
\label{sec:data-methods}
%=======================================================================

Our basic strategy in this study is to post-process the results of high-resolution simulations to produce synthetic \halpha\ spectral cubes (and cubes in some other, related lines), which we will then analyse using techniques chosen to match as closely as possible the standard procedures in observations. We begin with a brief review of the simulation data we use for this purpose in \autoref{ssec:qed_data}, and then describe our post-processing methods, including generating and fitting \halpha\ spectra (\autoref{sec:spectra} and \autoref{sec:fitting}, respectively), and deriving electron densities from density diagnostic lines (\autoref{sec:ne}). 

\subsection{QED III Data}
\label{ssec:qed_data}

\begin{table*}
\begin{center}
\begin{tabular}{cccccccccccc}

\hline\hline
Name & $\Sigma_{\rm SFR}$ & $\Delta x$ & $L_z$ & $\sigma_{\rm broad}$ & $\Sigma_{\rm{H}\alpha, \rm{broad}}$ & $t$ \\
 & [M$_\odot$ yr$^{-1}$ kpc$^{-2}$] & [pc] & [kpc] & [\kmps] & [$10^{37}$ $\mathrm{erg\,s^{-1}\,kpc^{-2}}$] & [Myr] \\
(1) & (2) & (3) & (4) & (5) & (6) & (7) \\
\hline
\multirow{5}{*}{$\Sigma$13-Z1-H150} & \multirow{5}{*}{$6\times10^{-3}$} & \multirow{5}{*}{2} & \multirow{5}{*}{4} & 25.42 & $9.18$ & 142 \\
 &  &  &  & 31.48 & $7.15$ & 160 \\
 &  &  &  & 42.13 & $7.88$ & 181 \\
 &  &  &  & 56.48 & $8.31$ & 210 \\
 &  &  &  & 43.08 & $12.0$ & 221 \\
\hline
\multirow{4}{*}{$\Sigma$13-Z0.2-H150} & \multirow{4}{*}{$6\times10^{-3}$} & \multirow{4}{*}{2} & \multirow{4}{*}{4} & 20.97 & $15.2$ & 230 \\
 &  &  &  & 14.13 & $1.70$ & 250 \\
 &  &  &  & 25.32 & $3.40$ & 270 \\
 &  &  &  & 23.76 & $3.14$ & 290 \\
\hline
\multirow{4}{*}{$\Sigma$2.5-Z1-H1500} & \multirow{4}{*}{$1.58\times10^{-4}$} & \multirow{4}{*}{4} & \multirow{4}{*}{8} & 18.03 & $0.0384$ & 153 \\
 &  &  &  & 19.93 & $0.0691$ & 200 \\
 &  &  &  & 20.96 & $0.0407$ & 249 \\
 &  &  &  & 23.38 & $0.0503$ & 299 \\
\hline
\hline
\end{tabular}
\caption{Simulation parameters for the three QED runs used in this work (including a subset of Table~1 in \qediii). Columns give: (1) run name; (2) star-formation surface density $\Sigma_{\rm SFR}$; (3) cell size $\Delta x$; (4) vertical box half-height $L_z$; (5) dispersion of the broad \halpha\ emission from the wing-only Gaussian fit, $\sigma_{\rm broad}$; (6) broad-component \halpha\ surface density $\Sigma_{\rm H\alpha, broad}$ (see \autoref{sec:fitting}); and (7) snapshot time $t$. Units are indicated in the header.}
\label{tab:params}
\end{center}
\end{table*}

\begin{figure*}
\centering
% \begin{subfigure}{0.49\textwidth}
%     \includegraphics[width=\textwidth]{figs/Outflow_Visualization_R8.pdf}
%     % \caption{$\rm \Sigma13-Z1-H150$ at $t=142$~Myr}
% \end{subfigure}
% \begin{subfigure}{0.49\textwidth}
%     \includegraphics[width=\textwidth]{figs/Outflow_Visualization_R8_0.2Zsol.pdf}
%     % \caption{$\rm \Sigma13-Z0.2-H150$ at $t=230$~Myr}
% \end{subfigure}
% \begin{subfigure}{\textwidth}
%     \includegraphics[width=\textwidth]{figs/Outflow_Visualization_R16_h1.5kpc_Zsol.pdf}
%     % \caption{$\rm \Sigma2.5-Z1-H1500$ at $t=153$~Myr}
% \end{subfigure}
\includegraphics[width=\textwidth]{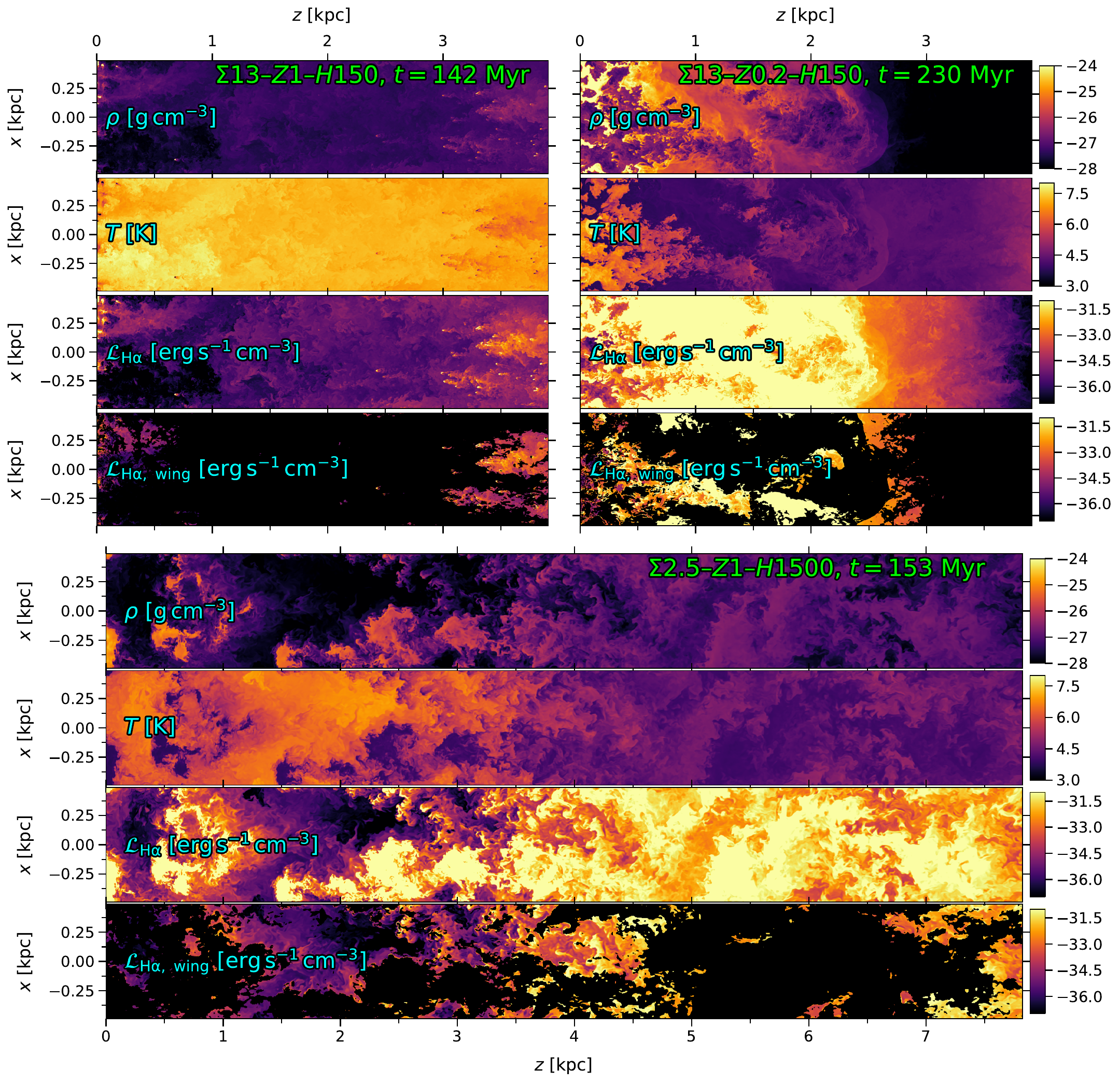}
\caption{Slices through the three simulation runs used in this study: (a) $\rm \Sigma13-Z1-H150$ at $t=142$~Myr, (b) $\rm \Sigma13-Z0.2-H150$ at $t=230$~Myr and (c) $\rm \Sigma2.5-Z1-H1500$ at $t=153$~Myr. Each panel shows, from top to bottom, the log gas density (in g cm$^{-3}$), log temperature (in K), log total \halpha\ emissivity (in erg s$^{-1}$ cm$^{-3}$), and log of the broad-wing ($50\le|v_z|/\mathrm{km\,s}^{-1}\le200$) \halpha\ emissivity (also in erg s$^{-1}$ cm$^{-3}$). Note that to enhance readability, we show only half of the simulation domain in these plots.}
\label{fig:vis}
\end{figure*}

In this study, we process the \citetalias{vijayan25} simulations, which comprise a set of ten simulations that follow outflows emerging from a 1 kpc$^2$ patch of a galactic disk; in all simulations, the galactic plane is centred at $z=0$, and the computational domain is a box extending from $(0, 0, -L_z)$ to $(L_x, L_y, +L_z)$ with $L_x = L_y = 1$ kpc. Simulations differ in their initial gas surface density and supernova rate (which are correlated), supernova scale height, metallicity (which affects the cooling rate), and vertical extent of the simulation domain $L_z$. Runs are denoted as $\Sigma$GG-Zzz-Hhhh, where `GG' indicates the initial gas surface density in M$_\odot$ pc$^{-2}$, `zz' indicates the gas-phase metallicity (normalised so that Solar metallicity corresponds to 1.0), and `hhh' indicates the scale height of supernova explosions in pc. Thus, for example, run $\rm \Sigma13-Z1-H150$ has an initial gas surface density of $13$ M$_{\odot}$ kpc$^{-2}$, Solar metallicity, and a supernova scale height of 150 pc. The complete set of QED runs is available in Table 1 of \citetalias{vijayan25}.

For this paper, we examine a subset of the full QED suite, which we summarise in \autoref{tab:params}. We focus on these cases because they produce outflows that contain significant quantities of the warm ($\sim 10^4$ K) gas that is the dominant source of \halpha\ emission; other QED simulations have outflows dominated by either cooler atomic and molecular material or hotter ($\sim 10^7$ K) supernova-shocked material. \autoref{fig:vis} shows slices of density and temperature for these three runs. For the colourbars we have chosen, warm gas embedded in the volume-filling hot medium is identifiable as dark patches in the temperature plots and corresponding bright yellow patches in the density plots. Warm gas that is present close to the midplane of the galaxy is part of the inter-stellar medium (ISM) of the galaxy, while that at higher altitudes results from a combination of entrainment and cooling.

In order to understand how the results may vary with time, even for a fixed galactic environment, due to the inevitable stochasticity of wind driving, we select four or five snapshots from each of our three sample simulations for analysis. Our goal is to sample (approximately) independent realisations of the outflow at fixed environment, rather than to follow the detailed evolution of a single outflow episode. We select our snapshots to be at times after the simulations have settled to statistical steady state (see \citetalias{vijayan25}), and to be well-enough separated in time for the outflow to have traversed much of the simulation domain, so that we are sampling independent realisations in each snapshot. This choice reduces correlations between adjacent snapshots and allows us to quantify the snapshot-to-snapshot scatter in the inferred H$\alpha$-based outflow properties. For this purpose, we select snapshots $\approx 20$ Myr apart in the case of \fid~and \subsol, and $\approx 50$ Myr apart for \outgal, which has a larger vertical extent. Note that our method is somewhat different from and complementary to approaches aimed at tracking the acceleration history of a single episode \citep[as sometimes adopted in absorption-line based studies; e.g.,][]{carr25}. The single episode approach is meaningful when there are distinct outflow episodes driven by, for example, external interactions. By contrast, simulations are intended to represent a statistically steady driving environment, though the instantaneous mass flux can be strongly time-variable in \subsol\ run. We list the times of the snapshots we select for analysis in \autoref{tab:params}.

\subsection{Generating \halpha\ spectra}
\label{sec:spectra}

We begin our analysis by computing the \halpha\ emissivity in each cell assuming case~B recombination.\footnote{In practice the \halpha~line is often blended with the [N~\textsc{ii}]6585 line in lower-resolution spectra. For simplicity we ignore this complication, and assume in our simulated observations that the contribution from [N~\textsc{ii}] can be removed. We therefore do not attempt to model [N~\textsc{ii}] emission.} Following \citet{draine11}, the effective recombination rate coefficient for gas of temperature $T$ is
\begin{equation}
\alpha_{\rm eff, H\alpha } \approx 1.17 \times 10^{-13} \ T_4^{-0.942-0.031 \,\, \mathrm{ln}(T_4)} \ \mathrm{cm^3 s^{-1}},
\end{equation}
with $T_4 \equiv T/10^4\,{\rm K}$. We estimate the local proton and electron number densities, $n_p$ and $n_e$, and the gas temperature $T$, from the \textsc{grackle} thermochemistry module used in the QED simulations \citep{smith17}. This module returns the temperature and mean molecular weight $\mu$ in each cell (using the total gas density and internal energy as inputs), which we convert to the abundances of free electrons and protons as follows. \textsc{Grackle} assumes a fixed ratio of 1 He per 10 H atoms and adopts a metal mass fraction $Z = 0.01295$ at Solar metallicity. The corresponding mass fractions of H, He, and metals are $(X,Y,Z) = (0.706, 0.281, 0.013)$ for the two runs with Solar metallicity (\fid\ and \outgal) and $(0.714, 0.284, 0.0026)$ for the run with 20\% solar metallicity (\subsol). The mass fractions are related to the number of free particles per unit mass (which is $\mu^{-1}$) as
\begin{equation}
    \mu^{-1} = (1 + x_e) X + \frac{Y}{4} + \frac{Z}{m_Z},
\end{equation}
where $\mu$ is in units of the proton mass, $x_e$ is the number of free electrons per H nucleon, $m_Z$ is the mean mass per atom for metal atoms, and we have neglected H$_2$ formation. We can then invert this relation to solve for $x_e$ in terms of $\mu$, and from this, we can deduce the number density of free electrons from the number density of H nucleons $n_\mathrm{H} = X\rho/m_p$ as
\begin{equation}
    n_e = x_e n_\mathrm{H} \approx \left[\left(\frac{1}{\mu}-\frac{Y}{4}\right)-1\right]\frac{\rho}{m_p},
\end{equation}
where, for the purposes of numerical evaluation, we have dropped the small term $Z/m_Z$. To obtain the number density of free protons, we assume that all electrons come from H if $x_e < 1$, and that H is fully ionized if $x_e > 1$, so that we have
\begin{equation}
    n_p = \min\left(\frac{X\rho}{m_p}, n_e\right).
\end{equation}
Given $n_e$, $n_p$, and $T$ for each cell in the simulation, the gas luminosity per unit volume in the \halpha\ line is
\begin{equation}
 \mathcal{L}_{\rm H\alpha } = \alpha _{\rm eff, H\alpha }h \nu_{\rm H\alpha } n_e n_p,
\label{eq:jHalpha}
\end{equation}
where $\nu_{\rm H\alpha}=457\,{\rm THz}$. We show example slices of \halpha\ luminosity ($L_{\rm{H}\alpha}$) computed via this procedure in the third rows of \autoref{fig:vis}.

To obtain a synthetic spectrum that would be seen by an observer looking face-on at the galaxy, in principle, we should convolve the \halpha\ emission in each cell with a line shape function representing the Gaussian distribution of particle velocities within the cell. However, since we are interested primarily in high-velocity components where the bulk velocities are large compared to the thermal dispersion, we simplify this procedure by treating the emission from each cell as a $\delta$-distribution that is all emitted at a frequency $\nu = \nu_\mathrm{H\alpha} (1 + v_z/c)$, where $v_z$ is the line-of-sight velocity for an observer located at $z = +\infty$. With this simplification, the observable emission per unit velocity per unit area in a velocity channel from $v_i$ to $v_i + \Delta v$ is
\begin{equation}
    \left(\frac{d\Sigma_{\rm H\alpha}}{dv}\right)_i = \frac{1}{A \Delta v} \sum_j \mathcal{L}_{\mathrm{H\alpha},j} V_j \Theta(v_{z,j}-v_i) \Theta(v_i + \Delta v-v_{z,j}),
    \label{eq:dsigma_dv}
\end{equation}
where $A = 1$ kpc$^2$ is the cross-sectional area of the simulation domain, $\mathcal{L}_{\mathrm{H\alpha},j}$, $V_j$, and $v_{z,j}$ as the \halpha\ luminosity per unit volume, volume, and $z$-velocity of cell $j$, $\Theta(x)$ is the Heaviside step function (zero for $x<0$, unity for $x>0$), and the sum runs over all cells $j$ of the simulation domain. In practice, we adopt $\Delta v = 5$ km s$^{-1}$ throughout this work. Note that in writing this expression, we have implicitly assumed negligible dust absorption. We make this assumption for simplicity, and to test the ability of \halpha\ emission to trace outflows under the best possible circumstances, when there are no complications from dust. 

\subsection{\halpha-inferred and true outflow rates}
\label{sec:fitting}

\begin{figure}
    \centering
    \includegraphics[width=\linewidth]{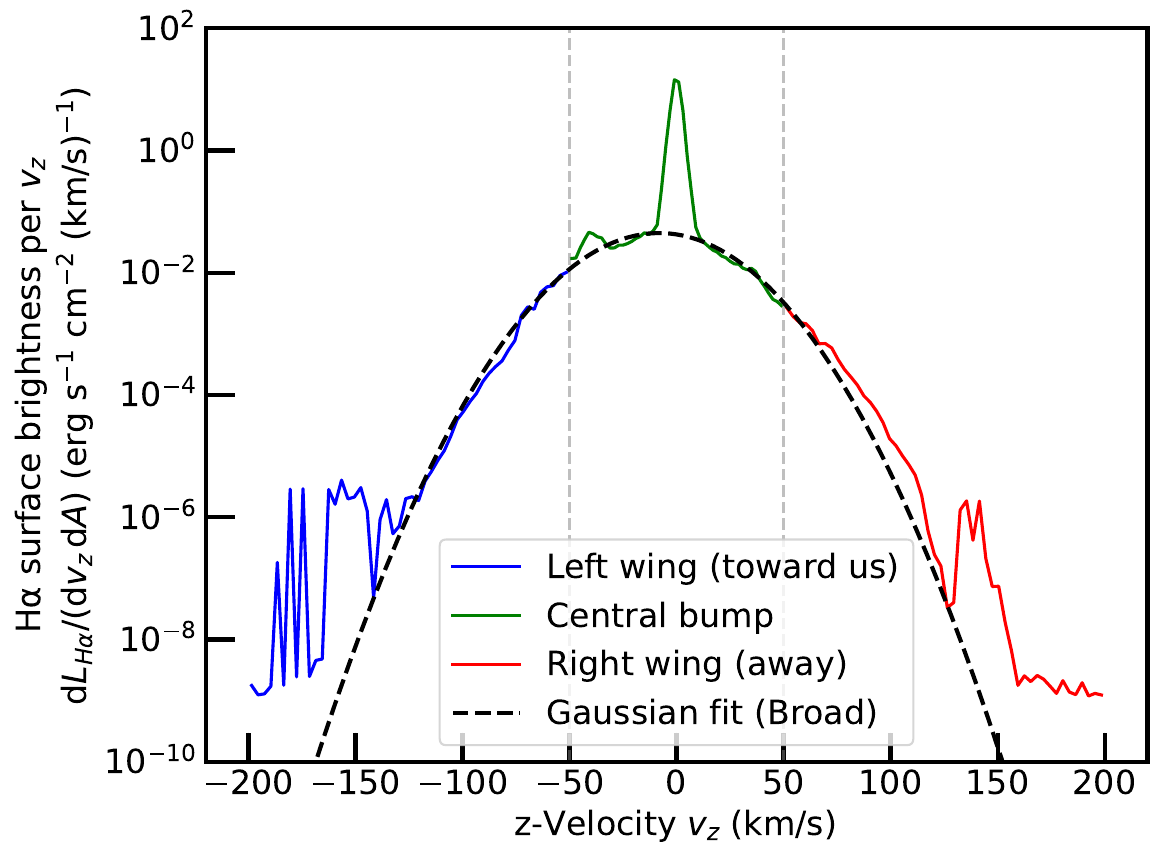}
    \caption{Example synthetic \halpha\ spectrum from the fiducial run $\Sigma13$--Z1--H150 at $t=142$ Myr (panel (a) from \autoref{fig:vis}). The dashed vertical lines at $v_z = \pm 50$ km s$^{-1}$ indicate the boundary between the wing region (red/blue solid lines) that we use to fit the broad line profile and the central narrow bump (green solid line) that we mask when fitting. The black dashed curve is the best-fitting Gaussian model to the broad component. We report the dispersion $\sigma_{\rm broad}$ and total integrated luminosity $\Sigma_{\mathrm{H}\alpha,\mathrm{broad}}$ of this fit in Columns 10 and 11 of \autoref{tab:params}. }
    \label{fig:halpha_fit}
\end{figure}

In \autoref{fig:halpha_fit}, we show a sample spectrum $d\Sigma_{\rm H\alpha}/dv$ we generate via the procedure described in \autoref{sec:spectra}; the example shown is for the 142 Myr snapshot from \fid\ run. We identify two distinct components: a central bump (solid, green line) that arises from low-velocity gas near the midplane of the galaxy, and broader components tracing warm gas accelerated in the outflows and flowing towards and away from the observer (highlighted in blue and red). Because we are interested in the latter, we carry out a least squares fit of a Gaussian profile to the broad wing components of the \halpha\ spectrum, where we define the broad wing as the emission in the velocity range $50 \leq |v_z|/\mathrm{km\,s}^{-1}\leq 200$; we show the emissivity in this velocity range in the bottom rows of the three panels in \autoref{fig:vis}. The dashed black line in \autoref{fig:halpha_fit} shows our fit to this example spectrum, and we carry out similar fits for all runs and snapshots. We report the dispersion $\sigma_\mathrm{broad}$ and the total, velocity-integrated \halpha\ luminosity per unit area $\Sigma_{\rm{H}\alpha, \rm{broad}}$ derived from each fit in \autoref{tab:params}. We have verified visually that Gaussian distributions provide reasonable fits to the spectral shape in all snapshots, and that using a lower threshold of $30$ \kmps\ to define the start of the broad wind changes our best-fitting values by less than 0.1 dex.

The quantities we extract from our fits -- $\Sigma_{\rm H\alpha,broad}$ and $\sigma_\mathrm{broad}$ -- are of interest because they are the critical quantities that enter into estimates of the mass outflow rate from the \halpha\ line. Here as an example we follow the calculation of the relationship between these quantities provided by \citet{reichardtchu22}, since their assumed planar geometry is well-matched to that of our simulations, but we note that the spherical versions of this argument typically adopted for more distant targets are conceptually similar. Let us imagine that the outflow consists of slabs of warm ionised material of constant electron and proton number densities $n_e$ and $n_p$ on either side of the galactic plane; these slabs extend from the disk to some height $H$ on either side of it, and move at constant velocity $v$ directly away from the disc. In this geometry, the total mass flux per unit area away from the disc is $\dot{\Sigma}_\mathrm{out} = 2 \mu m_\mathrm{H} n_p v$, where $\mu$ is the mean gas mass per H nucleon in amu ($\approx 1.4$ for Milky Way He abundances), and the factor of 2 comes from the two sides of the disk. The slabs will emit radiation in the \halpha\ line at a rate per unit volume given by \autoref{eq:jHalpha}, and thus the total \halpha\ luminosity per unit area emitted by the slabs is $\Sigma_{\rm H\alpha} = 2 \alpha_\mathrm{eff,H\alpha} h \nu_{\rm H\alpha} n_e n_p H$. If we identify this emission with that traced by the broad component of the \halpha\ line, and identify the velocity of the slab with the width of this broad emission, then we arrive at a relationship between the broad wing luminosity and the outflow rate in this simple slab model:
\begin{equation}
    \dot{\Sigma}_\mathrm{out,H\alpha} =  \frac{\mu m_\mathrm{H} \sigma_\mathrm{broad}}{\alpha_{\rm H\alpha,eff} h\nu_{\rm H\alpha} N_e} 
    \Sigma_\mathrm{H\alpha,broad}
    \label{eq: L_alpha}
\end{equation}
where $N_e = n_e H$ is the column density through one side of the slab. Thus for an assumed value of $N_e$ and our best-fit values of $\Sigma_{\rm H_\alpha,broad}$ and $\sigma_\mathrm{broad}$, we can derive an estimate of $\dot{\Sigma}_\mathrm{out,H\alpha}$ purely from observables, which we can compare to the true outflow rate measured from the simulations.

For clarity in what follows, we will always denote the H$\alpha$-derived outflow rate as $\dot{\Sigma}_\mathrm{out,H\alpha}$ in order to distinguish it from the true outflow rate, which we will denote $\dot{\Sigma}_\mathrm{out}$. Similar to Equation~4 in \qediii, we define the outflow rate at height $z$ as
\begin{equation}
    \dot{\Sigma}_{\rm out}(z) = \frac{1}{A}\left[\oint_{+z} \rho v_z \Theta(v_z) \, dx\, dy - \oint_{-z} \rho v_z \Theta(-v_z) \, dx\, dy\right],
    \label{eq:sigma_out}
\end{equation}
where the surface integrals are over the horizontal planes of thickness $dz=2$ pc parallel to the $xy$ plane located at heights $+z$ and $-z$ (where the disc midplane is at $z=0$). The area $A = 1$ kpc$^2$ is the full area of the simulation domain. Note here the Heaviside step functions $\Theta$, whose purpose is to select ``pure'' outward fluxes, i.e., we count only the mass with $v_z > 0$ on the $+z$ plane and $v_z < 0$ on the $-z$ plane. Thus, \autoref{eq:sigma_out} measures the instantaneous mass flux through the planes at $+z$ and $-z$ and sums the two sides. Tests using either the ``absolute'' flux (i.e., replacing the argument of the integrand with $\rho |v_z|$, so we count all mass passing through the surface in either direction) or the ``net'' flux (i.e., replacing the integrands with just $\rho v_z$, so we count the net flux through the surface) show noticeably less correlation with H$\alpha$ luminosity, confirming that the wing emission traces the instantaneous outward mass flux but carries no information on whether that material will escape or eventually rain back onto the disc. We discuss the implications of this further in \autoref{sec:limitations}.

% \avcomment{Column Blah of \autoref{tab:params} tabulates $\sigma$ values of the different runs. In Fig blah we show an example Halpha spectrum from blah run...}

% To explore the correlation between $L_{{\rm{H}\alpha}}$ and $\dot{M}$, we estimate mass outflow rates directly from simulations. $\dot{M}=\iint \rho v_z dA_{x,y}$ is evaluated on horizontal planes at heights $|z|=$ 0.5, 1, 2, 3, and 4 kpc, both above and below the midplane. In the runs we consider here, once outflows are established in a steady state, the mass outflow is positive, meaning the bulk velocity of the gas is outwards from the disc. However, the movement of gas is complex, and it is possible that some gas parcels are moving in, i.e. towards the disc at certain times. At any given time, within a plane in the simulation domain at a certain height from the disc, there may be gas cells with outward or inward $v_z$. To take into account the complexity of gas motions, we define the mass outflow rate in three different ways. ``Pure'' outflow rate is estimated by only using those cells for which $v_z$ points outwards from the disc, i.e, positive in the $z>0$ half and negative in the $z<0$ half. We estimate the ``net'' outflow rate by subtracting the inflow rate from the outflow rate. Lastly, to compute ``absolute'' outflow, we take the magnitude of $v_z$. 
% Additionally, every quantity is tabulated for both the full temperature range and a subset of warm gases. 

% To connect the simulated mass flux with the broad-wing luminosity, we invert the case-B recombination relation:

\subsection{Electron densities: [\ion{S}{ii}] doublet estimates and true values}
\label{sec:ne}

Use of \autoref{eq: L_alpha} requires knowledge of the electron column density $N_e$, which is generally not directly observable. One strategy that some authors have adopted to mitigate this uncertainty is to estimate the electron volume density in the outflow from density-sensitive doublet ratios, most prominently $[\ion{S}{ii}]\,\lambda\lambda6716,6731$ \citep[e.g.][]{Sanders16a, forsterSchreiber19, Weldon24a}. To evaluate how well this strategy works, for every simulation and snapshot time, we calculate two complementary estimates of the electron density: (i) a ``simulation truth'' $n_{e,\mathrm{true}}$ measured directly from the hydrodynamic state, and (ii) an observational estimate $n_{e,\ion{S}{ii}}$ obtained by inverting the $[\ion{S}{ii}]\,\lambda\lambda6716,6731$ doublet ratio. We evaluate both these quantities on thin horizontal slabs (thickness $\Delta z=60$ pc), and averaged over the $+z$ and $-z$ sides. This avoids mixing together emission from multiple heights and from the galactic disc, and such a clean separation would only be possible in a disc seen edge-on. Thus we are selecting the most favourable possible geometry for this method.

We define our true density as the \halpha~emissivity-weighted mean density, with the emissivity computed using \autoref{eq:jHalpha}. Quantitatively, we define
\begin{equation}
n_{e,\mathrm{true}}(z) =
 \frac{\int_{S(z)} n_{e} \mathcal{L}_{\rm H\alpha}\, dV} 
     {\int_{S(z)} \mathcal{L}_{\rm H\alpha} \, dV},
\label{eq:ne_true}
\end{equation}
% where $\mathcal{S}(|z|)$ denotes all cells within the two $\pm\,\Delta z/2$ slabs at the given $|z|$, and $\Delta V_i$ is the cell volume. 
where the volume of integration $S(z)$ is a pair of slabs on either side of the disc extending from $(z-30\,\mathrm{pc}, z+30\,\mathrm{pc})$ and $(-z-30\,\mathrm{pc}, -z+30\,\mathrm{pc})$. We perform this estimate for heights $z=\{0.5,1,2,3,4\}$ kpc. To obtain the corresponding densities from the [S~\textsc{ii}] doublet, for each cell we use \texttt{PyNeb} \citep{luridiana15} to compute the emissivities per unit volume in each of the two lines, which we denote $\epsilon_{6716}$ and $\epsilon_{6731}$, from the cell electron density and temperature, and assuming a fixed abundance of S$^+$ per proton; the choice of S$^+$ abundance does not matter for the purposes of computing the line ratio as long as it is uniform. We then integrate over the same slabs for which we have computed $n_{e,\rm true}(z)$ to obtain the total luminosity in each line emitted from that region:
\begin{equation}
L_{6713,6716}(z) = \int_{S(z)} \epsilon_{6713,6716}\, dV.
\end{equation}
We compute the slab–integrated ratio $R_{\ion{S}{ii}}\equiv L_{6716}/L_{6731}$, and use it to infer the density $n_{e,\rm S~\textsc{ii}}$ using the \texttt{getTemDen} routine from \texttt{PyNeb}, assuming a fixed diagnostic temperature $T=10^4$ K. Because the two lines are closely spaced in wavelength, $R_{\ion{S}{ii}}$ is effectively insensitive to reddening and thus we do not make any attempt to model the effects of dust. In some cases the ratio lies near (or outside) the low– or high–density asymptotes, and as a result the inversion is poorly constrained. We treat these cases as providing only upper or lower limits on the density, and omit them from density–based scalings; we note where we have applied this limits in the relevant figures below.

% [\ion{S}{ii}] lines diagnostic: I think instead of cloudy, there is another Python package looks useful, pyneb. 
% Checkout: https://pypi.org/project/pyneb/, 
% or the tutorial: https://github.com/Morisset/PyNeb_devel/blob/master/docs/Notebooks/PyNeb_manual_2.ipynb

%=======================================================================
\section{Results}
\label{sec:results}

Now that we have computed the H$\alpha$ and [S~\textsc{ii}] emission from all snapshots, we are in a position to answer our key questions about the use of H$\alpha$ as a diagnostic of galactic winds. We begin in \autoref{sec:phase_distribution_lalpha} with a qualitative examination of which gas, in terms of both location and properties, drives the H$\alpha$ wing emission. Then in \autoref{sec:correlation} we examine how well the properties of the broad H$\alpha$ emission correlate with true outflow rates, and we use this finding to propose a new, empirical calibration for deriving outflow rates from H$\alpha$ in \autoref{sec:Ne}. Finally, we examine the extent to which this calibration can be improved by the addition of the [S~\textsc{ii}] diagnostic in \autoref{sec:sii}.

\subsection{Origin of broad \halpha\ emission}
\label{sec:phase_distribution_lalpha} 

\begin{figure*}
\centering
$$
\begin{array}{c}
\includegraphics[width=\textwidth]{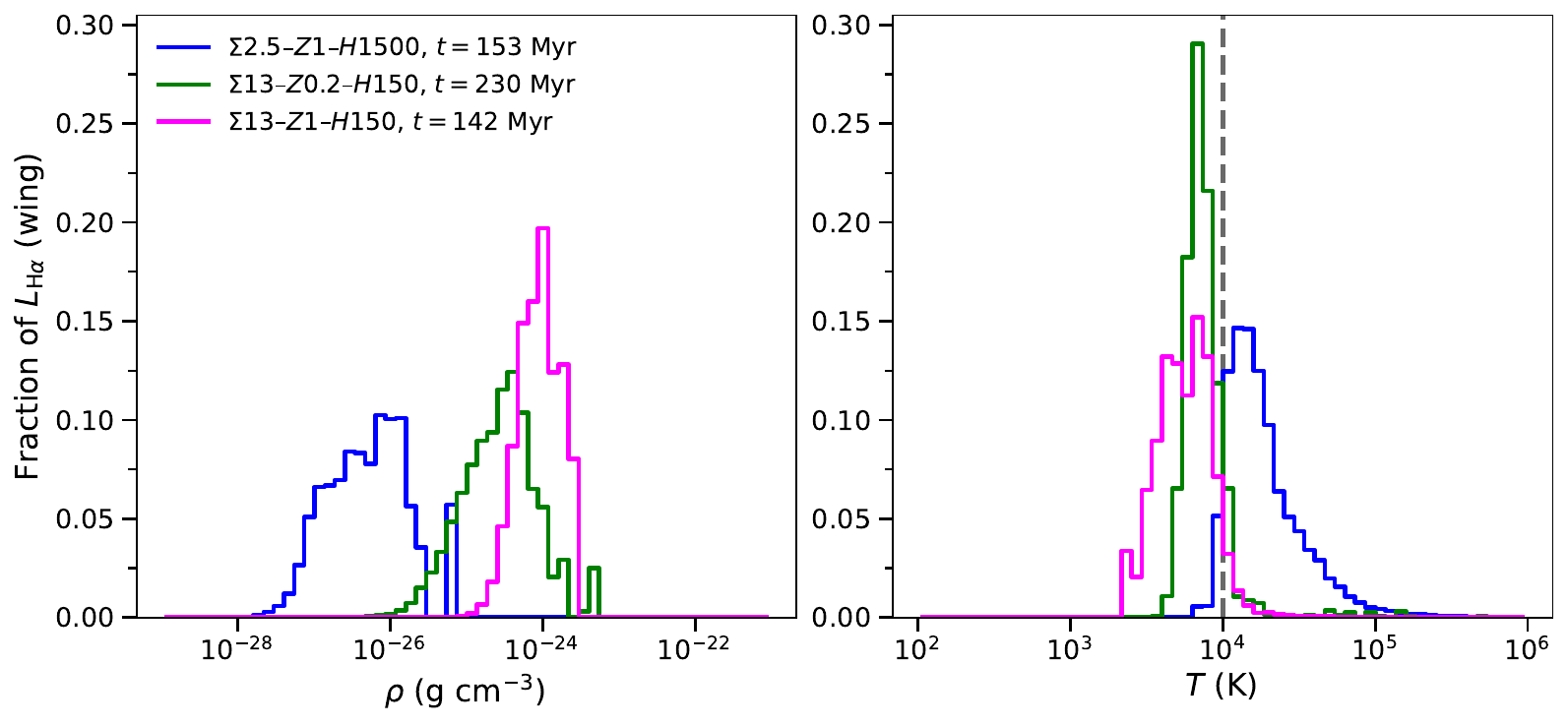}
\end{array}{}
$$
\caption{Left and right columns show the histograms of density and temperature, respectively, weighted by \halpha\ wing luminosity (i.e., luminosity emitted at velocities $50 \leq |v_z|/\mathrm{km\,s}^{-1}\leq 200$). Colours denote same snapshots as \autoref{fig:vis}: magenta is $\Sigma13$–$Z1$–$H150$ at $t=142$ Myr, green is $\Sigma13$–$Z0.2$–$H150$ at $t=230$ Myr and blue is $\Sigma2.5$–$Z1$–$H1500$ at $t=153$ Myr. The vertical dashed line marks $T=10^4$ K. }
\label{fig:hists}
\end{figure*}

% From \autoref{fig:vis}, we note that high-velocity \halpha\ emission comes from gas occupying a wide range in density and temperature, i.e., from different phases. We quantify the phase-distribution in \autoref{fig:hists}, which shows the luminosity–weighted distributions of density (left) and temperature (right) from the gas responsible for producing both the ``narrow'' ($|v_z| < 50$ km s$^{-1}$; blue) and ``wing'' (|$|v_z| > 50$ km s$^{-1}$; orange) parts of the spectrum generated from the same timesteps shown in \autoref{fig:vis}. For the \fid\ run, the wing emission arises from that that is systematically warmer and more diffuse than the narrower component that has begun to mix with the underlying hot wind. 

% For \subsol\ case, the wing and narrow emission come from overlapping temperature and density bins, following from the physically overlapping regions for them in \autoref{fig:vis}. Similarly for \outgal, while not entirely mutually exclusive, the distributions for narrow and wing spectrum are distinct, again reflecting the spatially distinct emission regions. 

We focus here on the origin of the broad \halpha\ component. \autoref{fig:hists} shows that the wing-emitting gas in the expected temperature range, $\sim 10^4$ K, but is much more diffuse, $\lesssim10^{-24}\ \mathrm{g\,cm^{-3}}$ (corresponding to $n_e \lesssim 1$ cm$^{-3}$ for fully ionzied gas), than is typically found in bright H~\textsc{ii} regions in the disc. There is also notable variation between runs, with the dwarf galaxy-like case, $\Sigma$2.5-Z1-H1500, showing the lowest density and highest temperature, while the Solar metallicity, Solar neighbourhood-like case, $\Sigma$13-Z1-H150, shows the highest density and lowest temperature.

% -------------------------------------------------
% \subsection{Origin of the \halpha\ wing emission}
% \label{sec:cdf}

\begin{figure}
\centering
$$
\begin{array}{c}
\includegraphics[width=0.49\textwidth]{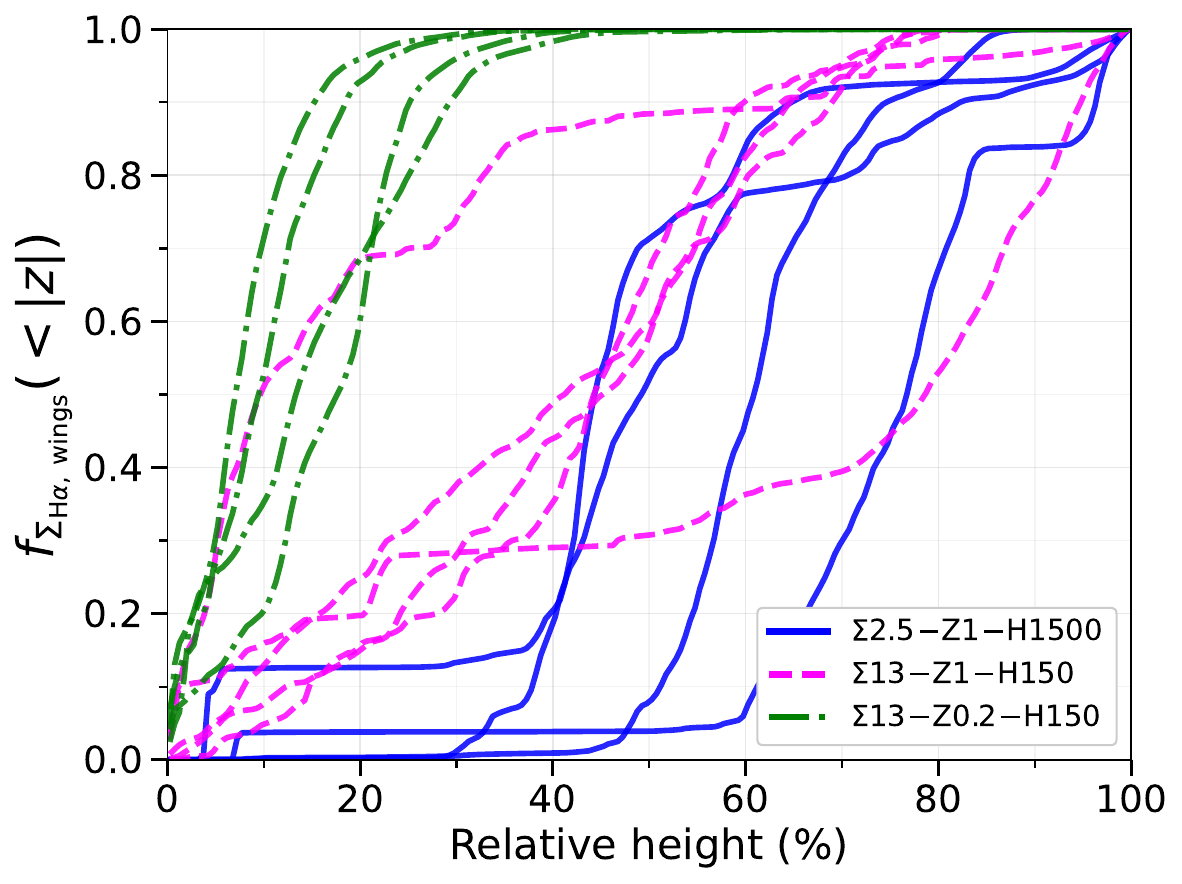}
\end{array}{}
$$
\caption{Cumulative fraction of the \halpha\ wing luminosity originating below a given $|z|$. Green, magenta and blue curves correspond to the $\rm \Sigma13-Z1-H150$, $\rm \Sigma13-Z0.2-H150$ and $\rm \Sigma2.5-Z1-H1500$ runs, respectively. Note here that the horizontal axis is expressed in units of the simulation domain half-height, which is 4 kpc for the two $\Sigma13$ runs but 8 kpc for $\rm \Sigma2.5-Z1-H1500$.}
\label{fig:cdf}
\end{figure}

To understand exactly where the \halpha\ wings are produced, we plot the cumulative distribution of \halpha\ wing emission in \autoref{fig:cdf}. We defined this quantity as the fraction of all \halpha~emission in the simulation volume that is emitted by cells within a distance $z$ of the midplane at $z=0$. The three simulation runs reveal significantly different distributions, which can be traced back to the kind of winds hosted by the galaxy. In the subsolar-metallicity run, \subsol\ (green), the cumulative fraction rises rapidly with $|z|$, indicating that the majority of the wing luminosity originates close to the disc. In contrast, the Solar-metallicity \fid\ (magenta) and the dwarf-like \outgal\ (blue) accumulate more gradually with height and show stronger snapshot-to-snapshot variation, implying a more extended and more time-variable origin for the wing-emitting gas. This behaviour is consistent with \autoref{fig:hists}, which shows that the wing-emitting gas in the dwarf-like run is weighted toward lower densities (and slightly higher temperatures) than in the $\Sigma$13 runs. % Bursty winds, observed in \subsol\ (green) and \outgal\ (blue) have a shallower increase in their wing emission. These runs are characterised by a large gas scale height, which inhibits sustained outflows. Supernova feedback injects hot gas near the midplane, which finds few low-density channels for effective escape. Consequently, warm gas is volume filling, creating a shallow profile for \halpha\ emission. On the other hand, the $\rm \Sigma13-Z1-H150$ run (magenta) hosts multiphase winds in which the hot outflowing winds entrain warm ISM material on their way out of the galaxy. The entrained gas exchanges mass with the surrounding hot medium and is continuously heated. This aligns with the fact that the wing luminosity, as shown in \autoref{fig:hists}, comes from hotter and less dense gas in this case than in the other two cases. The entrained clouds stop emitting once they are heated to a temperature too high for efficient hydrogen recombination. 

% This entrained material is the primary source of emission beyond disc scale height and it is continuously accelerated as evidenced by the rising contribution to mass loading from the cool ($T<2\times 10^4$ K) and warm components (see Figure 5 of \qediii). At about $\sim 1.5$ kpc most of this material has been heated out \halpha\ emission. 

These results demonstrate that the spatial origin of \halpha\ wing emission can vary by several kpc depending on how and from where in the disc of the galaxy the outflows are launched. Moreover, the vertical distribution shows high stochasticity in time even for a fixed set of galactic parameters (gas surface density, metallicity, star formation rate), with the median height from where emission arises fluctuating by factors of 2 or more over time. This high variability occurs because the \halpha\ wing luminosity is often dominated by a few bright, dense gas structures whose vertical positions fluctuate significantly over time (as is visible in \autoref{fig:vis}). For instance, a single bright clump of cool, dense gas near the simulation boundary can drive large excursions in the apparent emission height, as evidenced by the outlying dashed curves in \autoref{fig:cdf}. This stochasticity underscores that assuming a single characteristic wind height is unrealistic; instead, the \halpha\ wing emission traces the complex, time-variable distribution of dense gas structures within the multiphase outflow.

% -------------------------------------------------
\subsection{Correlation between broad H$\alpha$ emission and outflow rate}
\label{sec:correlation}

\begin{figure}
\centering
$$
\begin{array}{c}
\includegraphics[width=\columnwidth]{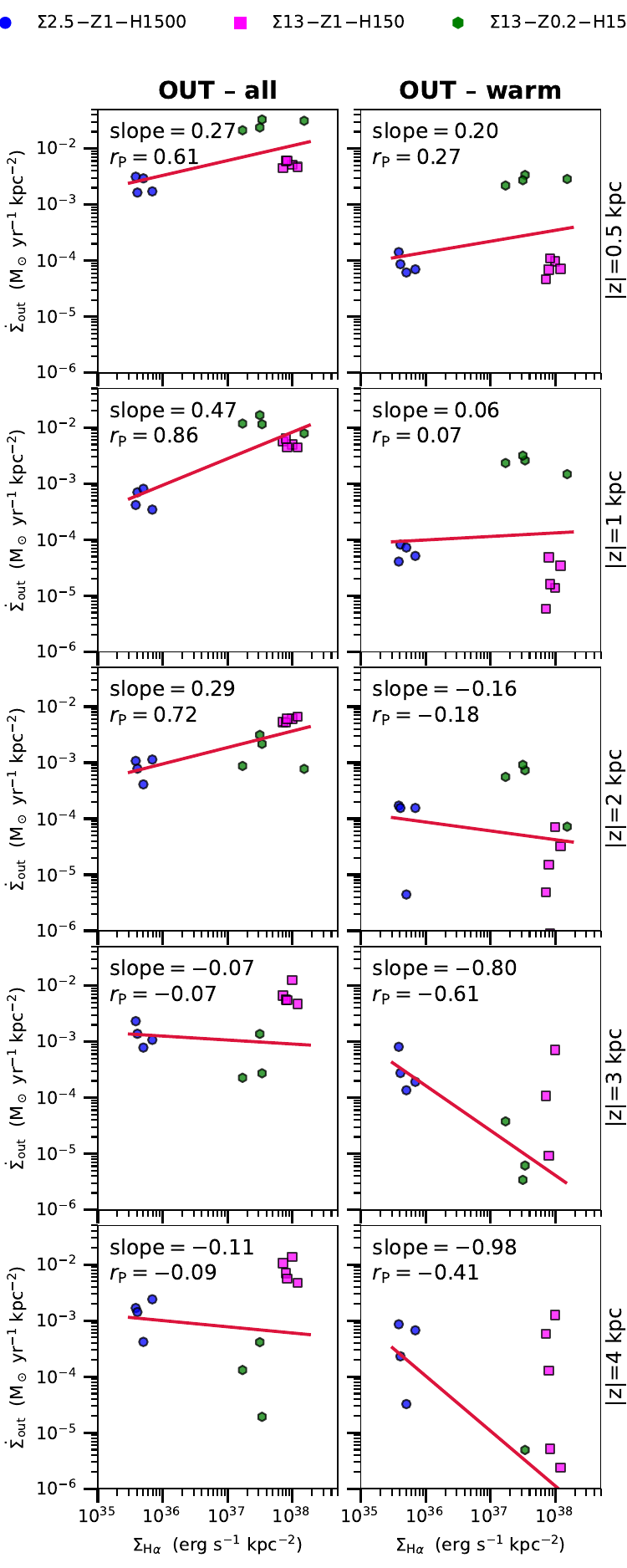}
\end{array}{}
$$
\caption{Relation between \halpha\ broad component surface brightness and mass-outflow rate surface density for all simulations (see legend).  Rows show different distances $|z|$ from the galactic plane, as indicated by the labels to the right, while the left and right columns show outflow rates for all gas (left) and only for ``warm'' gas at temperatures $1 \le T/10^4\,\mathrm{K} \le 3$ (right). Best-fit slopes and Pearson correlation coefficients $r_\mathrm{P}$ are indicated in each panel.}
\label{fig:mdot}
\end{figure}

From the previous section, we conclude that \halpha\ broad component emission originates from regions that vary with galactic environment, and over time even within a single environment. In order to investigate how well broad H$\alpha$ correlates with the mass outflow rate, we compare these two quantities in 
\autoref{fig:mdot}, evaluating $\dot{\Sigma}_\mathrm{out}$ at heights $z = 0.5$, 1, 2, and 4 kpc using \autoref{eq:sigma_out}. In the left column of this figure we plot the total mass outflow rate, while in the right column we isolate the outflow rate only considering warm gas, $10^{4} \le T/K \le 3\times10^{4}$, which one expect might correlate more closely with the H$\alpha$.
% The upper row considers gas at all temperatures, whereas the lower row restricts only to the warm phase ($10^{4} \le T/K \le 3\times10^{4}$).
We indicate the Pearson correlation coefficient and slope of the best-fit (OLS fit) line on the top left of each panel.
% somewhat arbitrary since the emission is never zero even from gas much cooler or hotter than $10^4$ K. The solid red lines show the best fit and the numbers indicate the best fit slope and the Pearson correlation coefficient.  

% Overall, we find an unsurprising positive correlation between the wing luminosity and the mass outflow rate. The correlation is stronger for all gas compared to warm which is expected since with a choice of the temperature range we might be missing \halpha\ wing emission by restricting the temperature (see Figure 2). 

The standard method for deriving outflow rate from broad H$\alpha$ emission under the assumption of a constant electron density and thickness of the outflow column, \autoref{eq: L_alpha}, implies that for fixed $N_e$ and $\sigma_\mathrm{broad}$, $\dot{\Sigma}_\mathrm{out}$ and $\Sigma_\mathrm{H\alpha,broad}$ should be linearly correlated. The figure shows that at lower heights, $|z|\lesssim 2$ kpc, the true outflow rate for all gas $\dot{\Sigma}_{\rm out}$ is indeed reasonably-well correlated with the surface brightness of the broad H$\alpha$ component, with Pearson correlations ranging from $\approx 0.6 - 0.9$. That the correlation is best at small heights is not surprising since $60-70\%$ of the total broad component emission arises from these heights at most times and runs (see \autoref{fig:hists}). However, {the best-fitting power-law index relating $\dot{\Sigma}{\rm out}$ and $\Sigma_\mathrm{H\alpha,broad}$ in log--log space is significantly below unity,} a point to which we will return in \autoref{sec:Ne}. A fit for the relation between $\dot{\Sigma}_{\rm out}$ and $\Sigma_\mathrm{H\alpha,broad}$ at $|z| = 1$ kpc, where the correlation is strongest, gives
\begin{equation}
\dot{\Sigma}_{\rm out}=0.022\times\left(\frac{\Sigma_{\rm{H}\alpha, \rm broad}}{10^{40} \,\mathrm{erg\, s^{-1}\,kpc^{-2}}}\right)^{0.47}M_\odot\,\mathrm{yr}^{-1}\,\mathrm{kpc}^{-2}, 
\label{eq: Mdot}
\end{equation}
with a Pearson coefficient $r_{\rm P} = 0.86$ computed for $\log \dot{\Sigma}_{\rm out}$ versus $\log \Sigma_{\rm H\alpha,broad}$, and at other heights $\leq 2$ kpc the slope of the correlation lies in the range $\approx 0.3-0.5$. 

We also note that, while naively we would expect to see a stronger correlation between the outflow rate of the warm gas and the broad component of H$\alpha$ emission, this naive assumption is false: warm outflow rate correlates with $\Sigma_\mathrm{H\alpha,broad}$ significantly more poorly than total outflow rate. We can understand this somewhat counter-intuitive result by examining the phase structure of the outflows as shown in \citetalias{vijayan24} and \citetalias{vijayan25}. A key finding of these papers is that, while warm ionised gas dominates the \textit{density} of the outflow at all heights, it only begins to contribute significantly to the \textit{mass flux} in the outflow once the gas reaches $\approx 3-4$ kpc off the plane, because the hot wind requires this distance to accelerate the warm gas up to full speed. In regions closer to the disc where the warm gas density and thus H$\alpha$ emission are strongest, much of this material is not yet moving fast enough to contribute to either the mass flux or the broad H$\alpha$ component. Its emission instead becomes lost within the narrow component associated with the disc. This explains the relatively poor correlation between wing emission and the warm gas mass flux. On the other hand, the more powerful the outflow, the more rapidly it is able to accelerate the warm gas up to speeds where it can contribute to the wing. This makes wing emission a reasonable proxy for the total flux in the outflow.
\subsection{Calibration for the electron column density}
\label{sec:Ne}

The sub-linear scaling between $\Sigma_{\rm{H}\alpha, \rm broad}$ and $\dot{\Sigma}_{\rm out}$ identified in \autoref{sec:correlation} suggests that the electron column density ($N_e \equiv n_e H$) may vary systematically with outflow strength, contradicting the standard observational assumption of constant $n_e$ and $H$ across galaxies with varying outflow properties. To quantify this variation and provide a practical correction, we invert \autoref{eq: L_alpha} and calculate the electron column density required to reproduce the simulated \halpha\ using the values of $\dot{\Sigma}_\mathrm{out}$ from the simulations at $|z|=1$ kpc, where the correlation is best. This requires that we adopt a value of $\sigma_\mathrm{broad}$, for which purpose we use the values returned by our fits to the H$\alpha$ wing emission (as reported in \autoref{tab:params}). In what follows we do not attempt to infer the physical, area-averaged electron column density of the outflow. Instead, we define an effective (or required) electron column density: the value that must be adopted in the \halpha\ estimator (\autoref{eq: L_alpha}) to reproduce the simulation-measured outflow rate at a given height. This effective $N_e$ implicitly accounts for the unresolved density inhomogeneity (e.g. clumping and multiphase structure) that enters recombination emission through its density weighting, and therefore should not be interpreted as a direct estimate of the true electron column \citep[cf.,][]{jennings25}.

\begin{figure}
\includegraphics[width=\columnwidth]{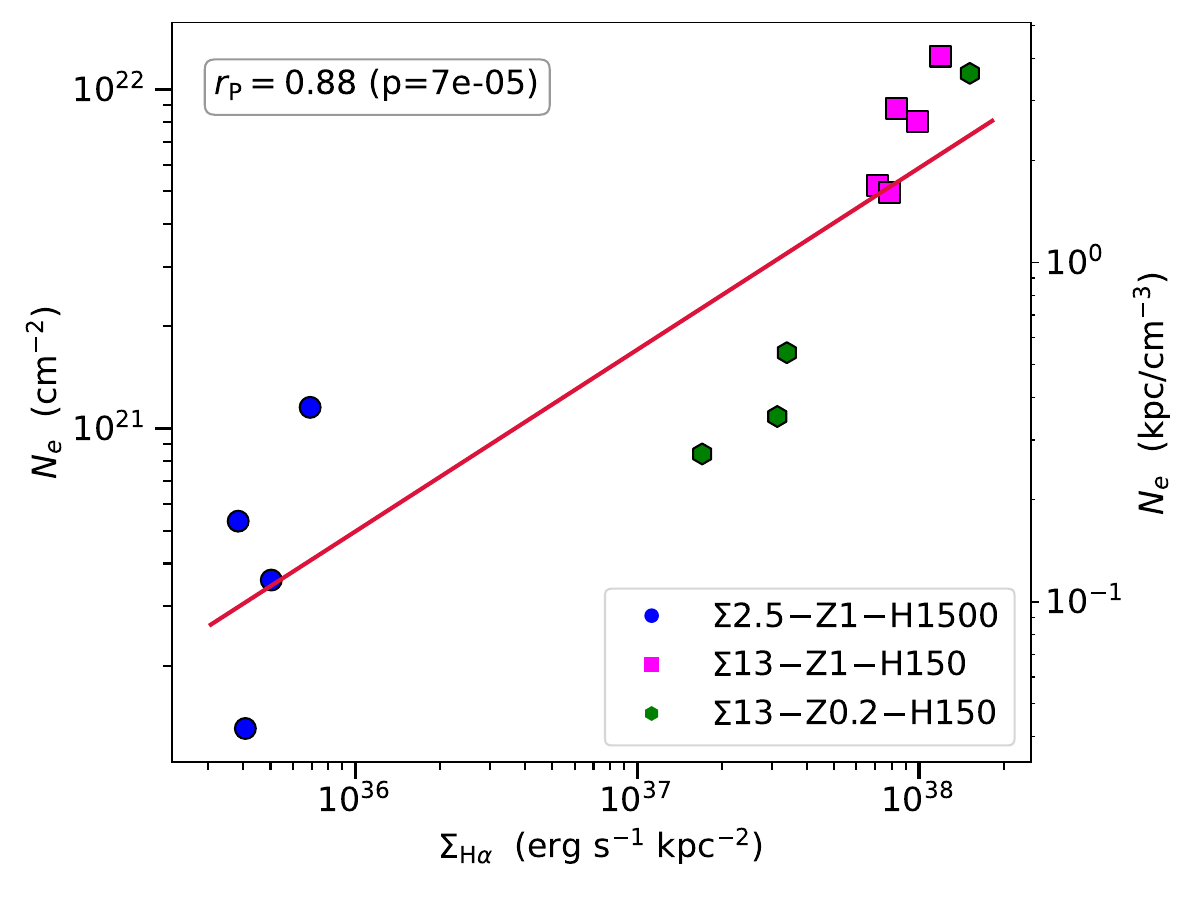}
\caption{Electron column density $N_e\equiv n_e H$ required to recover the true outflow rate at $|z| = 1$ kpc versus broad component \halpha\ surface brightness (see \autoref{eq: Ne}).}
\label{fig:Ne}
\end{figure}

\autoref{fig:Ne} shows the value of effective $N_e$ required to obtain the correct outflow rate at $|z| = 1$ kpc as a function of the broad component surface brightness $\Sigma_{\rm{H}\alpha, \rm broad}$. However, the results are qualitatively similar for any height $\lesssim 2$ kpc. The relationship exhibits a tight correlation with a Pearson coefficient $r_{\rm P} = 0.88$, and an OLS fit to the data gives a relation
\begin{equation}
N_e=N_{e,40}\left(\frac{\Sigma_{\rm{H}\alpha, \rm broad}}{10^{40}\,\mathrm{erg\,s^{-1}\,kpc^{-2}}}\right)^{0.54}
\label{eq: Ne}
\end{equation}
with
\begin{equation}
    N_{e,40} = 1.0\times 10^{23}\, \rm{cm}^{-2} = 33 \,\mathrm{kpc}\,\mathrm{cm}^{-3}.
\end{equation}

The tight, positive correlation between $N_e$ and $\Sigma_{\rm{H}\alpha, \rm broad}$ demonstrates that higher $\rm{H}\alpha$ broad emission, as expected for stronger outflows, is associated with systematically higher electron column densities. The power-law index of 0.54 here (i.e. the fitted slope in the $\log$--$\log$ space) is comparable to the slope found for the $\dot{\Sigma}_{\rm out}$--$\Sigma_{\rm{H}\alpha, \rm broad}$ relation, confirming that the electron density plays a crucial role in determining the observed \halpha\ emission. The practical implication is that observers can use \autoref{eq: Ne} to estimate the appropriate electron column density to adopt in \autoref{eq: L_alpha} for a given \halpha\ broad component measurement, thereby removing the systematic bias introduced by assuming a constant $n_e$. This numerical calibration provides a more physically motivated approach to converting \halpha\ broad component luminosities into mass outflow rates, accounting for the varying density structure of galactic winds across different outflow strengths. While \autoref{eq: Ne} is obtained by inverting \autoref{eq: L_alpha} using the simulation-measured $\dot{\Sigma}_{\rm out}$, its utility is demonstrated by the forward test in \autoref{sec:sii}, where adopting \autoref{eq: Ne} substantially improves recovery of $\dot{\Sigma}_{\rm out}$ compared to other common approaches in observation. Note that \autoref{eq: Ne} is a numerical calibration derived under the idealised assumptions listed in \autoref{sec:limitations}, and should be applied to observations with appropriate caution.

\subsection{Electron densities from the [\ion{S}{ii}] diagnostic}
\label{sec:sii}

\begin{figure}
\centering
\includegraphics[width=0.49\textwidth]{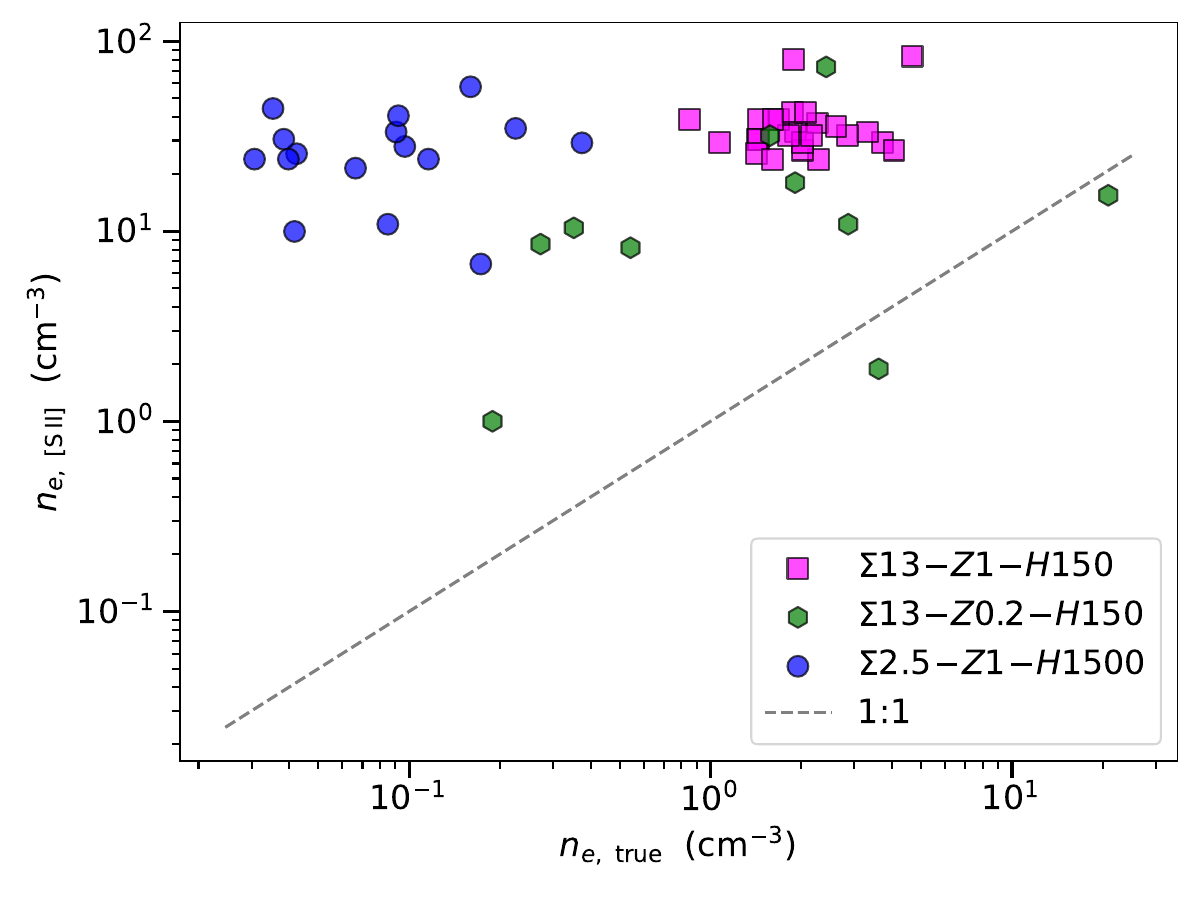}
\caption{Comparison of electron density $n_e$ derived from the [\ion{S}{ii}] doublet with the true electron density from the simulations. The dashed line indicates the one-to-one relation.}
\label{fig:sii_ne_comparison}
\end{figure}

% A popular diagnostic for inferring $n_e$ is via the [\ion{S}{ii}] doublet.  To illustrate how electron densities inferred from the [\ion{S}{ii}] doublet compare with the simulation truth, and to test the accuracy of the resulting mass outflow rate estimates, we include the following diagnostics. 

We have seen that estimates of the outflow rate derived from H$\alpha$ emission under the assumption of constant electron column density lead to significant errors, because the electron column density is in fact correlated with $\Sigma_\mathrm{H\alpha,broad}$. We next investigate whether these errors can be reduced by using a density diagnostic such as the [S~\textsc{ii}] doublet. We compare the true (i.e., H$\alpha$ emission-weighted using \autoref{eq:ne_true}) and [\ion{S}{ii}]-based electron densities in \autoref{fig:sii_ne_comparison}. Evidently, [\ion{S}{ii}] is a biased probe most sensitive around $\sim 10$ cm$^{-3}$. This is expected as the [\ion{S}{ii}] line ratio is less sensitive to electron density in the low-density regime \citep[$n_e \lesssim 100$ cm$^{-3}$; see][for review]{kewley19}. However, our true electron densities frequently lie outside this range, meaning the [S~\textsc{ii}] significantly overestimates them.

\begin{figure}
\centering
$$
\begin{array}{c}
\includegraphics[width=\columnwidth]{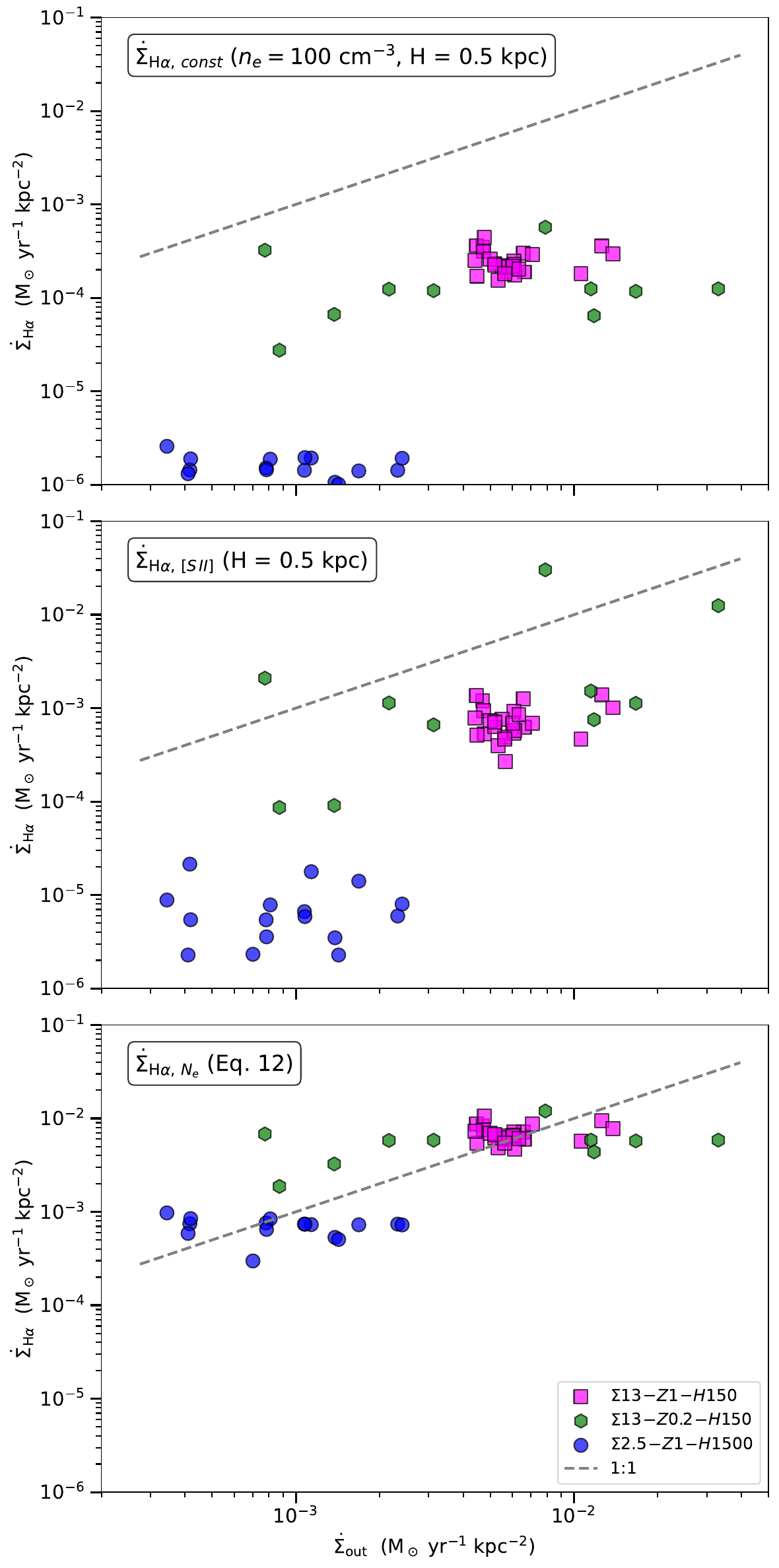}
\end{array}{}
$$
\caption{Top: Comparison of mass outflow rates derived using a fixed electron density of $100\,{\rm cm^{-3}}$ with the true mass outflow rates from simulations. Middle: Comparison of mass outflow rates derived using [\ion{S}{ii}]-based densities with the true mass outflow rates from simulations. Bottom: Comparison of mass outflow rates derived using our effective $N_e$ calibration \autoref{eq: Ne} with the true mass outflow rates from simulations. The dashed lines indicate the one-to-one relation.}
\label{fig:mdot_comparison}
\end{figure}

To test the effect of using [S~\textsc{ii}]-based electron density estimates, we compute an outflow rate $\dot{\Sigma}_{\rm out,H\alpha-\ion{S}{ii}}$ from \autoref{eq: L_alpha} using $n_{e,\ion{S}{ii}}$ for the electron density and a constant height $H = 0.5$ kpc. We compare this to outflow rates estimated using the same $H$ with constant $n_e = 100$ cm$^{-3}$, and to those using our empirical calibration (\autoref{eq: Ne}) at $|z|=[0.5, 1, 2, 3,4]$ kpc in \autoref{fig:mdot_comparison}. From the figure, it is clear that both the [\ion{S}{ii}]-based and constant $n_e$ assumptions lead to a systematic underestimate of the mass outflow rate, one that worsens at small outflow rates; this is another manifestation of the sublinear correlation between $\dot{\Sigma}_{\rm out}$ and $\Sigma_{\rm H\alpha,broad}$ we found in \autoref{sec:correlation}. Moreover, the [\ion{S}{ii}]-based estimate is only slightly superior to simply the assumption of constant $n_e$, and performs significantly worse than our empirical calibration. This may at least in part because the range of electron densities present in our simulations is too low for the [\ion{S}{ii}] diagnostic to be effective; it is possible that [\ion{S}{ii}] might perform better for starburst galaxies with denser outflows. We return to this point in \autoref{sec:limitations}. For non-starbursting galaxies of the type targeted by, for example, MaNGA \citep{rodriguez19, Avery+21} and SAMI \citep{zovaro24}, however, [\ion{S}{ii}] appears to be at most marginally useful as an additional constraint.

%This bias arises because \halpha\ luminosity per unit ionised mass decreases in the low-density regime, and because the [\ion{S}{ii}] doublet becomes increasingly insensitive to electron density for $n_e \lesssim 100 {\rm cm^{-3}}$ \citep{kewley19}, causing the [\ion{S}{ii}]-inferred $n_e$ to be poorly constrained or biased. As a result, using [\ion{S}{ii}]-based densities does not fully eliminate the systematic underestimation of $\dot{M}$ relative to the ground truth from simulation, even though it provides locally motivated density estimates. 

\section{Discussion}
\label{sec:disscusion}

Here we discuss the implications of our findings for observations (\autoref{subsec:obs_comparison}) and compare our results to those from previous theoretical work (\autoref{subsec:sims_comparison}). We conclude with some caveats and directions for future work (\autoref{sec:limitations}).

\subsection{Implications for the interpretation of observations}
\label{subsec:obs_comparison}

As discussed in \autoref{Sec:intro}, many \halpha-based outflow studies convert broad component luminosity to outflow rate assuming typical electron densities $n_e\!\sim\!80$–$400~{\rm cm^{-3}}$  \citep{newman12,genzel14,fiore17,reichardtchu22,reichardtchu25}. In some cases these densities (and the corresponding length scales, which are equally important but less-often discussed) are derived from density estimators such as [\ion{S}{ii}], while in others they are simply assumed values. For example, \citet{reichardtchu22} derive pixel-by-pixel mass outflow rates assuming $n_e \simeq 380$ cm$^{-3}$, which is an average over a sample of galaxies that spans two orders of magnitude variation in inferred mass outflow rate \citep{Schreiber+19}. However, our findings suggest that assuming a constant $n_e$ for galaxies that span a large range of outflow properties is problematic because we find that the effective electron column $N_e\propto \Sigma_{\rm H\alpha}^{0.54}$, so $N_e$ co-varies with outflow strength. Such systematic trends introduce a bias in the inferred scaling of outflow rate with other galaxy properties. In particular, adopting a single canonical $N_e$ tends to overestimate the steepness of variation of outflow rate with other galaxy properties. Thus for example \citet{reichardtchu25} report $\dot{\Sigma}_\mathrm{out} \propto \dot{\Sigma}_\mathrm{SFR}^{1.08}$, where $\dot{\Sigma}_\mathrm{SFR}$ is the star formation rate per unit area, but this is based on converting H$\alpha$ to outflow rate using fixed $N_e$. If our QED-based result that $\dot{\Sigma}_\mathrm{out} \propto \Sigma_{\rm H\alpha,broad}^{0.47}$ (driven by $N_e\propto \Sigma_{\rm H\alpha,broad}^{0.54}$) applies to the physical conditions probed by their sample, then a fixed-$N_e$ conversion would systematically steepen the inferred $\dot{\Sigma}_\mathrm{out}$–$\dot{\Sigma}_\mathrm{SFR}$ scaling. As an order-of-magnitude illustration, applying this sub-linear H$\alpha$–outflow scaling would map a slope of $1.08$ to an effective slope of $\sim 1.08\times 0.47 \approx 0.5$. We acknowledge that quantitative applicability of this re-interpretation is limited by the parameter space covered by our simulations and by the assumptions discussed in \autoref{sec:limitations}.
% Thus for example \citet{reichardtchu22} report $\dot{\Sigma}_\mathrm{out} \propto \dot{\Sigma}_\mathrm{SFR}^{1.07}$, where $\dot{\Sigma}_\mathrm{SFR}$ is the star formation rate per unit area, but this is based on converting H$\alpha$ to outflow rate using fixed $N_e$; our finding that outflow rate only varies as $\Sigma_{\rm H\alpha,broad}^{0.47}$ suggests that true relation is significantly flatter, closer to $\dot{\Sigma}_\mathrm{out} \propto \dot{\Sigma}_\mathrm{SFR}^{0.5}$.

Our finding that [\ion{S}{ii}] is of limited use in improving these estimates, at least for the types of outflows sampled in our simulations, also has implications for observations. They suggest that for surveys such as SAMI and MaNGA that target non-starburst galaxies, this strategy is of limited use, and calls into question whether the strategy is applicable even in galaxies with denser outflows  \citep[e.g.,][]{watts24}. Our recommendation is therefore to treat $[\ion{S}{ii}]$–based electron densities as an upper limit, and complement these upper limits with our $N_e(\Sigma_{{\rm H}\alpha,{\rm broad}})$ calibration, particularly for low-luminosity or dim outflows. This finding should also motivate additional work on trans–auroral line ratios (e.g. [\ion{S}{ii}]$\ \lambda\lambda$4068,4076 / $\lambda\lambda$6716,6731; [\ion{O}{ii}]$\ \lambda\lambda$7319,7330) that can work at lower densities. Unfortunately this remain observationally demanding and thus to this point have mostly been used in AGN samples \citep{spence18,rose18}.

\subsection{Comparison with previous simulations}
\label{subsec:sims_comparison}

% Our findings are consistent with stratified-disk simulations that emphasise how vertical structure governs whether the warm phase is entrained and at what heights it accelerates. In TIGRESS, a hot wind escapes with roughly constant flux above $\sim$kpc heights, whereas the warm component largely forms a fountain with declining flux at larger $|z|$; burstiness reflects correlated SF/SN activity, while radiative cooling limits hot–warm coupling \citep{kim18,kim20,vijayan20}. What QED adds is an observable calibration in \halpha\: a slope and normalisation for $\dot{\Sigma}_{\rm out}$ versus $\Sigma_{{\rm H}\alpha,{\rm broad}}$ (Equation~\ref{eq: Mdot}), and an empirical $N_e(\Sigma_{{\rm H}\alpha,{\rm broad}})$ that reconciles the sub–linear scaling.

Though there are numerous simulations in the literature exploring galactic outflow properties, and a reasonable number carrying out simulated observations in the X-ray regime \citep[e.g.,][]{Schneider24a, huang25} or for absorption lines \citep[e.g.,][]{peeples19, acharyya25}, fewer have also carried out simulated observations of the H$\alpha$ emission line. In part this is due to technical challenges: \halpha\ emissivity scales as $n_e n_p$, and as a result much of the emission is produced in dense clouds or in the mixing layers around them (as we have seen in \autoref{fig:vis}), which are difficult to capture without $\sim$pc-scale resolution. Lagrangian methods, which focus resolution in dense material at the price of having lower resolution in warmer, more diffuse material, find capturing H$\alpha$ particularly hard. Despite this, some authors have carried out synthetic H$\alpha$ observations prior to this work.

One notable example is \citet{ceverino16}, who compute simulated \halpha\ emission from a set of redshift $\sim 2$ zoom-in cosmological runs with high star formation rates. They find narrow and broad components in their \halpha\ spectra, similar to those seen in our work and in observations, though their median values for the width of the broad component is $95$ \kmps, much higher than ours. They posit that recent mergers in galaxies could lead to broader \halpha\ profiles. It is also possible that the higher widths are simply due to the much higher star formation rates in their simulated galaxies compared to ours, which target local rather than $z\sim 2$ conditions. Interestingly, however, \citeauthor{ceverino16} find fountain flows to be dominant in their galaxies up to a height of $\sim 2$ kpc, much the same as in our simulations.
% with synthetic \halpha\ at $z{\sim}2$ reproduce two-Gaussian line profiles and ubiquitous outflows; however, high-velocity gas is often confined to thick discs/fountains at small radii because of deep central potentials \citep{ceverino16}. This mirrors our finding that the $\Sigma_{{\rm H}\alpha,{\rm broad}}$–$\dot{\Sigma}_{\rm out}$ link is tightest near $|z|\!\sim\!1$ kpc and weakens at larger heights as fountain mixing and \halpha-dark phases grow (Section~\ref{sec:cdf}). 

More recently, \citet{howatson25} carried out simulations of winds driven by isolated disc galaxies simulated using FIRE-2 subgrid physics \citep{Hopkins18b} coupled to the \textsc{chimes} chemistry module \citep{Richings14a, Richings14b}. They use these simulations to predict a range of wind observables, including H$\alpha$ and [\ion{S}{ii}]. Compared to our simulations, theirs have the advantage of capturing the full disc and thus avoiding some of the limitations arising from our tall box geometry that we discuss in \autoref{sec:limitations}. However, this comes at the price of significantly lower resolution and therefore more reliance on subgrid physics, in comparison to the QED simulations that fully resolve the Sedov-Taylor phase and thus do not require a subgrid module. Despite these differences, their results are consistent with our conclusion that electron densities are by far the largest uncertainty in converting H$\alpha$ emission to outflow rates, that $[\ion{S}{ii}]$-based electron densities can overestimate true values by up to $\sim$2 dex for true electron densities  $n_e\lesssim 1$ cm$^{-3}$. This supports our conclusion that \halpha-only inferences require a density calibration like that provided by \autoref{eq: Ne} to avoid large systematics.

% Finally, there is a mature ecosystem for hot-phase X-ray synthesis and instrument convolution \citep[e.g., \texttt{pyXSIM} and \texttt{SOXS};][]{zuhone16,zuhone23} that has been used alongside global wind suites such as CCGOLS to generate mock soft X-ray images and spectra \citep{schneider18, schneider20}. By contrast, there are comparatively fewer studies that synthesise \halpha\ emission from resolved outflows. The bottlenecks are well-known: \halpha\ emissivity scales as $\propto n_e n_p$, which makes predictions strongly resolution dependent; much of the emission is produced in thin and turbulent mixing layers that require parsec-scale grids and non-equilibrium ionisation. Thus, a large fraction of the synthetic observation studies has focused on absorption-line diagnostics of the CGM and winds, as established by the FOGGIE simulations \citep{peeples19, acharyya25}. In this context, QED provides resolved \halpha\ emission measurements and derives scaling relations linking the broad component \halpha\ surface brightness to outflow properties. 

\subsection{Limitations and Future Work}
\label{sec:limitations}

Several limitations of our current analysis point toward important directions for future work. 
% First, our simulations span a limited range of galactic environments and do not include AGN feedback, which may drive outflows with different density structures. Extending the QED suite to include AGN-driven outflows would test whether our calibrations apply to the most luminous systems observed at high redshift. 
First, the finite size of our simulation domain ($\pm 4-8$ kpc) allows significant amount of fast-moving gas to escape the box even though it would have contributed to the \halpha\ emission in a real galaxy. Further, \qediii\ imposes a diode boundary condition which prevents gas from entering the simulation domain. As a result, we are unable to capture the H$\alpha$ emission that would be expected to arise from fountain gas that falls back from larger heights. A full disc simulation is required to alleviate this issue.
% Larger simulation domains or cosmological zoom-in simulations would be needed to distinguish between material that escapes and material that returns as part of the galactic fountain. This distinction is crucial for connecting \halpha\ observations to long-term galaxy evolution. 
Second, the current generation of QED simulations use a star formation rate that is prescribed rather than computed self-consistently from the gas dynamics. This is potentially problematic when it comes to comparing our simulations to observations correlating the properties of the \halpha~spectrum, for example its width or the number of components present within it, with galaxy star formation properties \citep[e.g.,][]{rodriguez19, Avery+21, zovaro24}.

Third, our synthetic spectra assume perfect velocity resolution and do not account for observational effects such as instrumental broadening, atmospheric seeing, or finite signal-to-noise ratios. For example, the SAMI and MaNGA surveys report a velocity resolution of $26$ and $70$ \kmps, respectively, which are comparable to and larger than the width of our broad component \citep{Croom+12, Bundy+15}. More realistic forward modelling that includes these effects would provide additional guidance for interpreting real observations. 

Fourth, our synthetic emission modelling and line decomposition are intentionally idealised. We assume case~B recombination, neglect dust attenuation and scattering, and adopt a face-on viewing geometry; we also define the ``broad'' emission using a fixed velocity window ($50 \le |v_z|/\mathrm{km,s}^{-1}\le 200$) and characterise it with a single wing-only Gaussian fit. These assumptions are designed to test \halpha\ as an outflow diagnostic under controlled (best-case) conditions, but they also mean that our $N_e(\Sigma_{\rm H\alpha,broad})$ relation (i.e., \autoref{eq: Ne}) should be interpreted as a numerical calibration for this setup rather than a universal observational prescription. In particular, applying \autoref{eq: Ne} to observational data will require accounting for dust, inclination-dependent mixing of disc and wind emission, and survey-specific spectral resolution and S/N, all of which can bias the recovered $\Sigma_{\rm H\alpha,broad}$ and $\sigma_{\rm broad}$.

Lastly, the QED simulation suite focuses on conditions in nearby galaxies. As a result, it does not presently include galaxies with the types of high star formation rates commonly found in main sequence galaxies as cosmic noon, or local galaxies selected to be their analogues. This means that our results are well-suited to provide calibrations to local galaxy surveys such as SAMI and MaNGA, but do not cover the range of parameters typically found in moderate- to high-redshift applications of \halpha-derived outflow rates. Even for local galaxies, it seems likely that our simulations do not cover the full range of possible outflow conditions. For example, the typical broad component width reported by \citet{zovaro24} is $\sim 86$ \kmps, which is larger than our fitted values of $\sigma_{\rm broad}$ (\autoref{tab:params}), although by less than a standard deviation ($42$ \kmps in the SAMI sample); the mean in the SAMI sample may also be biased upwards by the fact that broad components narrower than $\sim 40$ \kmps are undetectable due to SAMI's velocity resolution. Nonetheless, it is clear that at least some local galaxies have \halpha~wings broader than our simulations produce. This may be a result of the limited range of environments we sample, but it may also reflect the limitations of our tall box geometry, which means that there is no orbital motion contribution to the outflow velocity as might happen in a real galaxy seen at an angle that is not perfectly face-on.

% Finally, we have focused exclusively on \halpha\ emission, but other diagnostic lines (e.g., [O III], [N II]) probe different ionization states and density regimes. A comprehensive multi-line analysis could provide additional constraints on outflow properties and test the generality of our electron density calibration.

\section{Conclusion}
\label{sec:conclusion}

We have used high-resolution simulations of galactic patches from the QED simulation suite \citep{vijayan24, vijayan25} to investigate the relationship between broad \halpha\ emission and galactic outflows, covering a range of gas content and metallicity from Solar-neighbourhood-like to dwarf-like. Our major conclusions are as follows:

\begin{enumerate}

    \item \textbf{Broad wing emission arises from a wide range of conditions and locations, depending on the environment and varying over time.} The gas emitting \halpha\ in the high-velocity part of the spectrum is relatively diffuse (see \autoref{fig:hists}), and largely consists of material at the boundaries between a hot outflow and the denser, cooler material it is entraining (\autoref{fig:vis}). The median height above the galactic plane from which this emission comes varies by factors of $\sim 2$ or more over time, even for galactic patches of fixed gas content and star formation rate. This variation is a result of the stochastic motion of dense clouds or other features as they are entrained into the wind. 
    
    \item \textbf{Mass outflow rates within $\approx 2$ kpc of the galactic plane are correlated with the broad  component of the \halpha\ spectrum, but the correlation disappears at larger heights.} The correlation is strongest at a height of $|z|=1$ kpc, where we find $\dot{\Sigma}_{\rm out}\propto(\Sigma_{\rm{H}\alpha, \rm broad})^{0.47}$ with Pearson coefficient $r_{\rm P} = 0.86$ (see \autoref{eq: Mdot}). The disappearance of the correlation at larger heights is due to fountain flows, which contribute to H$\alpha$ emission near the plane but then do not go on to yield outflow at larger heights. Nevertheless, the tight correlation we find at low to intermediate heights validates the use of broad \halpha\ emission as an outflow mass flux indicator in the near-disc region, with the important caveat that the quantity measured is the near-disc mass flux rather than the flux that ultimately escapes the galaxy.

    \item \textbf{While broad \halpha~does correlate well with near-disc outflow rates, the correlation is substantially sub-linear because the electron column density, $N_e$, also scales systematically with outflow rates.} We find that $N_e \propto (\Sigma_{\rm{H}\alpha, \rm broad})^{0.54}$ (see \autoref{eq: Ne}), meaning that stronger outflows are associated with higher electron column densities. This invalidates the widespread observational assumption of fixed $N_e$, and implies that correlations between outflow rates and other galaxy properties (e.g., star formation rates) derived using that assumption are likely to be significantly too steep compared to reality. We provide a numerical calibration for the electron column density as a function of broad H$\alpha$ surface brightness (\autoref{eq: Ne}) that can be used to remove this bias. This numerical calibration is derived under idealised conditions (no dust, case~B recombination, face-on geometry, and our wing-fitting procedure) with a limited parameter space and is intended as a controlled baseline for interpreting observations, not a universal prescription.

    \item \textbf{Electron density estimates from density diagnostic line ratios, most prominently  $[\ion{S}{ii}]\,\lambda\lambda6716,6731$, systematically overestimate the electron density in situations when the true density is low. Consequently, derivations of outflow rates using these derived electron densities only slightly outperform those derived assuming a fixed $N_e$.} Instead, both the assumption of a fixed $N_e$ and the use of an electron column assuming fixed thickness but using an $[\ion{S}{ii}]$-derived $n_e$ lead to significant overestimates of the density, and thus significant underestimates of the outflow rate. Thus even locally measured $[\ion{S}{ii}]$ densities cannot fully remove systematics in H$\alpha$-based outflow rates. A practical path forward is to treat $[\ion{S}{ii}]$ densities as upper limits in the low-density regime and to combine direct line-ratio constraints with our empirical column-density calibration, $N_e(\Sigma_{{\rm H}\alpha,{\rm broad}})$, using the latter to regularize the conversion wherever the doublet is insensitive.

\end{enumerate}

\section*{Software}

This research made use of \texttt{astropy} \citep[\url{https://www.astropy.org/}]{astropy22}, \texttt{numpy} \citep[\url{https://numpy.org}]{numpy}, \texttt{matplotlib} \citep[\url{https://matplotlib.org/}]{matplotlib}, \texttt{yt} \citep[\url{https://yt-project.org/}]{turk11}, \texttt{PyNeb} \citep[\url{https://pypi.org/project/pyneb/}]{luridiana15} and \textsc{Quokka} \citep[\url{https://github.com/quokka-astro/quokka}]{wibking22, He24a}.

\section*{Acknowledgements}

We thank the anonymous referee for a helpful review of this work. AV and MRK acknowledge support from the Australian Research Council through awards FL220100020 and DP230101055. The simulations suite QED and the mock data discussed in this paper were generated with the assistance of resources from the National Computational Infrastructure (NCI Australia), an NCRIS enabled capability supported by the Australian Government, and from the Pawsey Supercomputing Research Centre's Setonix Supercomputer (\url{https://doi.org/10.48569/18sb-8s43}, with funding from the Australian Government and the Government of Western Australia. AV would like to thank H.~R.~M.~Zovaro for valuable insights and discussions. 

\section*{Data Availability}

The software pipeline used in this analysis is available from \url{https://github.com/Rongjun-ANU/QEDIV}. Due to their large size, the raw QED simulation outputs on which the analysis is performed are not available in this repository, but are available on reasonable request to the authors.



\bibliographystyle{mnras}
\bibliography{reference} % if your bibtex file is called example.bib

%\begin{thebibliography}{99}
%\end{thebibliography}

%%%%%%%%%%%%%%%%%%%%%%%%%%%%%%%%%%%%%%%%%%%%%%%%%%

%%%%%%%%%%%%%%%%% APPENDICES %%%%%%%%%%%%%%%%%%%%%

\appendix

% \section{}

%%%%%%%%%%%%%%%%%%%%%%%%%%%%%%%%%%%%%%%%%%%%%%%%%%

% Don't change these lines
\bsp	% typesetting comment
\label{lastpage}
\end{document}